\documentclass[sigconf]{acmart}
\renewcommand\footnotetextcopyrightpermission[1]{} %
\usepackage{eurosym}
\usepackage{booktabs}
\usepackage{flushend}
\usepackage{graphicx}
\usepackage{subcaption}
\usepackage{enumitem}
\usepackage{url}
\usepackage{amssymb,amsmath}
\usepackage[T1]{fontenc}
\usepackage[hyphenbreaks]{breakurl}
\newcommand{\descr}[1]{\smallskip\noindent\textbf{#1}}

\makeatletter
\def\@copyrightspace{\relax}
\makeatother
\makeatletter
\def\@copyrightspace{\relax}
\makeatother
\makeatletter
\def\@copyrightspace{\relax}
\makeatother
\begin{document}   
\title{Mean Birds: Detecting Aggression and Bullying on Twitter}

\author{Despoina Chatzakou$^{\dagger}$, Nicolas Kourtellis$^{\ddagger}$, Jeremy Blackburn$^{\ddagger}$\\[-2ex]
Emiliano De Cristofaro$^{\sharp}$, Gianluca Stringhini$^{\sharp}$, Athena Vakali$^{\dagger}$}
\affiliation{$^\dagger$Aristotle University of Thessaloniki \hspace{0.2cm} $^\ddagger$Telefonica Research \hspace{0.2cm} $^{\sharp}$University College London\\
deppych@csd.auth.gr, nicolas.kourtellis@telefonica.com, jeremy.blackburn@telefonica.com\\
e.decristofaro@ucl.ac.uk, g.stringhini@ucl.ac.uk, avakali@csd.auth.gr}

\renewcommand{\shortauthors}{D. Chatzakou et al.}

\copyrightyear{2017}
\acmYear{2017}
\setcopyright{acmcopyright}
\acmConference{WebSci '17}{June 25-28, 2017}{Troy, NY, USA}
\acmPrice{15.00}\acmDOI{http://dx.doi.org/10.1145/3091478.3091487}
\acmISBN{978-1-4503-4896-6/17/06}

\newenvironment {squishlist}
{\begin{list}{$\bullet$}
  { \setlength{\itemsep}{1pt}
     \setlength{\parsep}{1pt}
     \setlength{\topsep}{1pt}
     \setlength{\partopsep}{1pt}
     \setlength{\leftmargin}{1.5em}
     \setlength{\labelwidth}{1em}
     \setlength{\labelsep}{0.5em} } }
{\end{list}}

\newcommand{\nknote}[1]{\textcolor{blue}{\small \bf [NK: #1]}}

\begin{abstract}
In recent years, bullying and aggression against social media users have grown significantly, causing serious consequences to victims of all demographics. Nowadays, cyberbullying affects more than half of young social media users worldwide, suffering from prolonged and/or coordinated digital harassment. Also, tools and technologies geared to understand and mitigate it are scarce and mostly ineffective. In this paper, we present a principled and scalable approach to detect bullying and aggressive behavior on Twitter. We propose a robust methodology for extracting text, user, and network-based attributes, studying the properties of bullies and aggressors, and what features distinguish them from regular users. We find that bullies post less, participate in fewer online communities, and are less popular than normal users. Aggressors are relatively popular and tend to include more negativity in their posts. We evaluate our methodology using a corpus of $1.6$M tweets posted over $3$ months, and show that machine learning classification algorithms can accurately detect users exhibiting bullying and aggressive behavior, with over $90\%$ AUC.
\end{abstract}

\maketitle

\section{Introduction}\label{sec:intro}
Cyberbullying and cyberaggression are serious and widespread issues affecting increasingly more Internet users. Arguably, in today's hyper-connected society, bullying, once limited to particular places or times of the day (e.g., school hours), can instead occur anytime, anywhere, with just a few taps on a keyboard.
Cyberbullying and cyberaggression can take many forms and definitions~\cite{grigg2010cyber,Smith2008CyberbullyingNature,
Tokunaga2010CriticalReviewCyberbullyingVictimization}, however, the former typically denotes repeated and hostile behavior performed by a group or an individual and the latter intentional harm delivered via electronic means to a person or a group of people who perceive such acts as offensive, derogatory, harmful, or unwanted~\cite{grigg2010cyber}.

Only a few years ago, when Internet adoption was still limited, cyberbullying was not taken seriously: the typical advice was to ``just turn off the screen'' or ``disconnect''~\cite{screenbullying1}. 
However, as its proliferation and the extent of its consequences reach epidemic levels~\cite{cyberbullyingResearchCenter}, this behavior can no longer be ignored: in 2014, about 50\% of young social media users reported being bullied online in various forms~\cite{factsaboutbullying}.
Popular social media platforms like Twitter and Facebook are not immune~\cite{salon}, as racist and sexist attacks may even have caused potential buyers of Twitter to balk~\cite{guardiantrolls}.

In this paper, we focus on these phenomena on Twitter. Despite the seriousness of the problem, there are very few successful efforts to detect abusive behavior on Twitter, both from the research community (see Section~\ref{sec:related-work}) and Twitter itself~\cite{guardianfail}, due to several inherent obstacles.
First, tweets are short and full of grammar and syntactic flaws, which makes it harder to use natural language processing tools to extract text-based attributes and characterize user interactions.
Second, each tweet provides fairly limited context, therefore, taken on its own, an aggressive tweet may be disregarded as normal text, whereas, read along with other tweets, either from the same user or in the context of aggressive behavior from multiple users, the same tweet could be characterized as bullying.
Third, despite extensive work on spam detection in social media~\cite{GiatsoglouCSFV15,stringhini2010detecting,Wang2010SpamDetectionTwitter}, Twitter is still full of spam accounts~\cite{Chen2015SpamTweets}, 
often using vulgar language and exhibiting behavior (repeated posts with similar content, mentions, or hashtags) that could also be considered as aggressive or bullying.
Filtering out such accounts from actual abusive users may be a difficult task.
Finally, aggression and bullying against an individual can be performed in several ways beyond just obviously abusive language -- e.g., via constant sarcasm, trolling, etc.

\descr{Overview \& Contributions.} 
In this paper, we design and execute a novel methodology geared to label aggressive and bullying behavior in Twitter.
Specifically, by presenting a principled and scalable approach for eliciting user, text, and network-based attributes of Twitter users, we extract a total of $30$ features. 
We study the properties of bullies and aggressors, and what features distinguish them from regular users, alongside labels provided by human annotators recruited from CrowdFlower~\cite{crowdflower}, a popular crowdsourcing platform.
Such labels, contributed infrequently or at regular intervals, can be used to enhance an already trained model, bootstrapping the detection method and executed on large set of tweets.

We experiment with a corpus of $1.6M$ tweets, collected over 3 months, finding that bully users are less ``popular'' and participate in fewer communities.
However, when they do become active, they post more frequently, and use more hashtags, URLs, etc., than others.
Moreover, we show that bully and aggressive users tend to attack, in short bursts, particular users or groups they target.
We also find that, although largely ignored in previous work, network-based attributes are actually the most effective features for detecting aggressive user behavior.
Our features can be fed to classification algorithms, such as Random Forests, to effectively detect bullying and aggressive users, achieving up to $0.907$ weighted AUC~\cite{hanley1982meaning}, $89.9\%$ precision, and $91.7\%$ recall.
Finally, we discuss the effectiveness of our methods by comparing results with the suspension and deletion of accounts as observed in the wild for users who, though aggressive, remain seemingly undetected by Twitter.
Our datasets are available to the research community upon request.
\vspace{-0.3cm}

\section{Related Work}\label{sec:related-work}
Over the past few years, several techniques have been proposed to measure and detect offensive or abusive content / behavior on platforms like Instagram~\cite{Hosseinmardi2015}, YouTube~\cite{Chen2012DetectingOffensiveLanguage}, 4chan~\cite{4chan}, Yahoo Finance~\cite{Djuric2015HateSpeechDetection}, and Yahoo Answers~\cite{kayes2015ya-abuse}.
Chen et al.~\cite{Chen2012DetectingOffensiveLanguage} use both textual and structural features (e.g., ratio of imperative sentences, adjective and adverbs as offensive words) to predict a user's aptitude in producing offensive content in YouTube comments, while Djuric et al.~\cite{Djuric2015HateSpeechDetection} rely on word embeddings to distinguish abusive comments on Yahoo Finance. 
Nobara et al.~\cite{Nobata2016AbusiveLanguageDetection} perform hate speech detection on Yahoo Finance and News data, using supervised learning classification. 
Kayes et al.~\cite{kayes2015ya-abuse} find that users tend to flag abusive content posted on Yahoo Answers in an overwhelmingly correct way (as confirmed by human annotators). Also, some users significantly deviate from community norms, posting a large amount of content that is flagged as abusive. Through careful feature extraction, they also show it is possible to use machine learning methods to predict which users will be suspended.

Dinakar et al.~\cite{Dinakar2011ModelingDetectionTextualCyberbullying} detect cyberbullying by decomposing it into detection of sensitive topics. 
They collect YouTube comments from controversial videos, use manual annotation to characterize them, and perform a bag-of-words driven text classification.
Hee et al.~\cite{Hee2015AutomaticDetectionPreventionCyberbullying} study linguistic characteristics in cyberbullying-related content extracted from Ask.fm, aiming to detect fine-grained types of cyberbullying, such as threats and insults.
Besides the victim and harasser, they also identify bystander-defenders and bystander-assistants, who support, respectively, the victim or the harasser.
Hosseinmardi et al.~\cite{Hosseinmardi2015} study images posted on Instagram and their associated comments to detect and distinguish between cyberaggression and cyberbullying.
Finally, authors in~\cite{Saravanaraj2016} present an approach for detecting bullying words in tweets, as well as demographics about bullies such as their age and gender.

Previous work often used features such as punctuation, URLs, part-of-speech, n-grams, Bag of Words (BoW), as well as lexical features relying on dictionaries of offensive words, and user-based features such as user's membership duration activity, number of friends/followers, etc.
Different supervised approaches have been used for detection:~\cite{Nobata2016AbusiveLanguageDetection} uses a regression model, whereas~\cite{Dadvar2014ExpertsMachinesAgainstBullies,Dinakar2011ModelingDetectionTextualCyberbullying,Hee2015AutomaticDetectionPreventionCyberbullying} rely on other methods like Naive Bayes, Support Vector Machines (SVM), and Decision Trees (J48).
By contrast, Hosseinmardi et al.~\cite{Hosseinmardi2014TowardsUnderstandingCyberbullying} use a graph-based approach based on likes and comments to build bipartite graphs and identify negative behavior. 
A similar graph-based approach is also used in~\cite{Hosseinmardi2015}. 

Sentiment analysis of text can also contribute useful features in detecting offensive or abusive content.
For instance, Nahar et al.~\cite{Nahar2012SentimentAnalysisDetectionCyberBullying} use sentiment scores of data collected from Kongregate (online gaming site), Slashdot, and MySpace.
They use a probabilistic sentiment analysis approach to distinguish between bullies and non-bullies, and rank the most influential users based on a predator-victim graph built from exchanged messages.
Xu et al.~\cite{Xu2012FastLearningSentimentAnalysisBullying} rely on sentiment to identify victims on Twitter who pose high risk to themselves or others.
Apart from using positive and negative sentiments, they consider specific emotions such as anger, embarrassment, and sadness.
Finally, Patch~\cite{patch2015detecting} studies the presence of such emotions (anger, sadness, fear) in bullying instances on Twitter.

\descr{Remarks.} Our work advances the state-of-art on cyberbullying and cyberaggression detection by proposing a scalable methodology for large-scale analysis and extraction of text, user, and network based features on Twitter, which has not been studied in this context before.
Our novel methodology analyzes users' tweets, individually and in groups, and extracts appropriate features connecting user behavior with a tendency of aggression or bullying.
We examine the importance of such attributes, and further advance the state-of-art by focusing on new network-related attributes that further distinguish Twitter-specific user behaviors.
Finally, we discuss the effectiveness of our detection method by comparing results with the suspension and deletion of accounts as observed in the wild for users who, though aggressive, remain seemingly undetected.

\section{Methodology}\label{sec:methodology}
\begin{figure}[!t]
	\centering
	\includegraphics[width=0.45\textwidth]{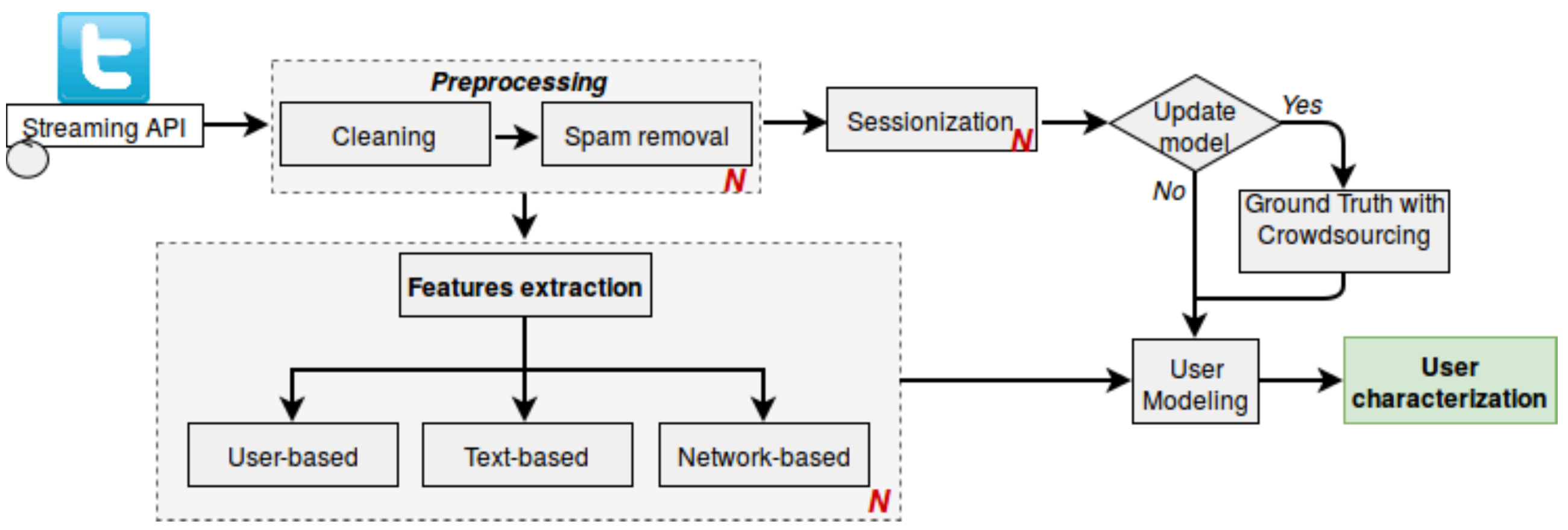}
	\vspace{-0.1cm}
	\caption{Overview of our methodology. {\em N} denotes the ability to parallelize a task on $N$ processors.}
	\label{fig:methodology}
	\vspace{-0.3cm}
\end{figure}

Our approach to detect aggressive and bullying behavior on
Twitter, as summarized in Figure~\ref{fig:methodology}, involves the following steps: (1) data collection, (2) preprocessing of tweets, (3) sessionization, (4) ground truth building, (5) extracting user-, text-, and network-level features, (6) user modeling and characterization, and (7) classification.

\descr{Data Collection.} Our first step is to collect tweets and, naturally, there are a few possible ways to do so.
In this paper, we rely on Twitter's Streaming API,  which provides free access to $1\%$ of all tweets. 
The API returns each tweet in a JSON format, with the content of the tweet, some metadata (e.g., creation time, whether it is a reply or a retweet, etc.), as well as information about the poster (e.g., username, followers, friends, number of total posted tweets).

\descr{Preprocessing.} Next, we remove stop words, URLs, and punctuation marks from the tweet text and perform normalization -- i.e., we eliminate repetitive characters; e.g., the word ``yessss'' is converted to ``yes''.
This step also involves the removal of spam content, which can be done using a few different techniques relying on tweeting behavior (e.g., many hashtags per tweet) and/or network features (e.g., spam accounts forming micro-clusters)~\cite{GiatsoglouCSFV15,Wang2010SpamDetectionTwitter}.

\descr{Sessionization.} Since analyzing single tweets does not provide enough context to discern if a user is behaving in an aggressive or bullying way, we group tweets from the same user, based on time clusters, into {\em sessions} and analyze them instead of single tweets.

\descr{Ground Truth.} We build ground truth (needed for machine learning classification, explained next) using human annotators. 
For this we use a crowdsourced approach, by recruiting workers who are provided with a set of tweets from a user, and are asked to classify them according to predefined labels.
If such an annotated dataset is already available, this step can be omitted.

\descr{Feature Extraction.} We extract features from both tweets and user profiles.
More specifically, features can be user-, text-, or network-based; e.g., the number of followers, tweets, hashtags, etc.

\descr{Classification.} The final step is to perform classification using the (extracted) features and the ground truth. 
Naturally, different machine learning techniques can be used for this task, including probabilistic classifiers (e.g., Na\"ive Bayes), decision trees (e.g., J48), ensembles (e.g., Random Forests), or neural networks. 

\descr{Scalability and Online Updates.} An important challenge to address is supporting {\em scalable} analysis of large tweet corpora.
Several of the above steps can be parallelized, e.g., over {\em N} subsets of the data, on {\em N} cores (Figure~\ref{fig:methodology}).
Also, one can use different modeling algorithms and processing platforms (e.g., batch platforms like Hadoop vs. stream processing engines like Storm) if data are processed in batches or in a streaming fashion.
Either way, some of the steps (e.g., annotation from crowd-workers) can be periodically executed on new data, and the model updated to handle changes in data and/or manifestation of new aggressive behaviors.
Our pipeline design provides several benefits with respect to performance, accuracy, and extensibility, as it allows regular updates of the model, thus capturing previously unseen human behaviors. 
Moreover, we can plug-in new features, e.g., as new metadata is made available from the Twitter platform, or from new research insights. 
Finally, different components can be updated or extended with new technologies for better data cleaning, feature extraction, and modeling.

\section{Dataset  \& Ground truth} \label{sec:data-ground-truth}

In this section, we present the data used in our evaluation, and the way we process it to build ground truth.

\begin{figure}[!t]
	\centering
	\begin{subfigure}[b]{0.21\textwidth}
		\includegraphics[width=\textwidth]{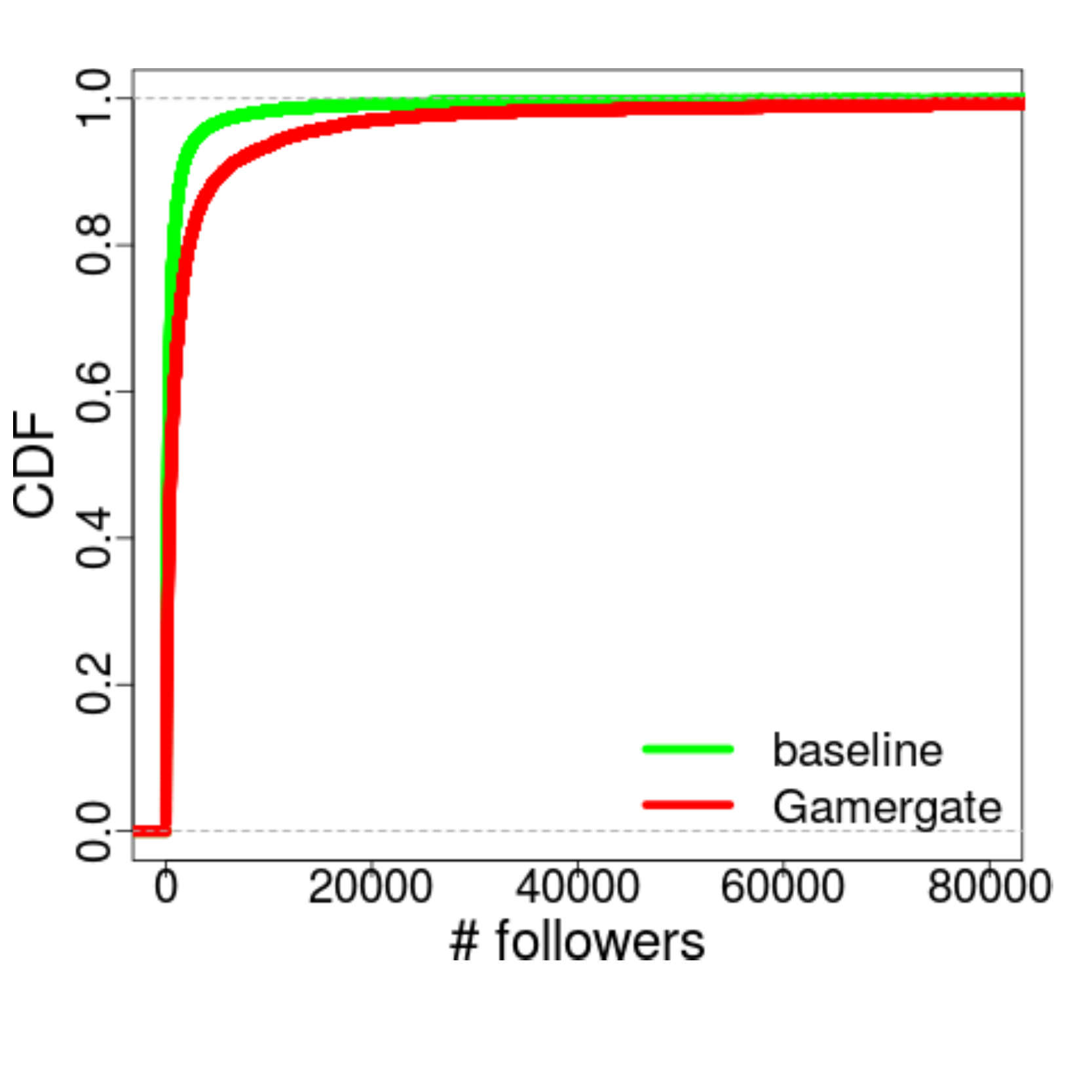}
		\captionsetup{font=scriptsize}
		\vspace{-0.85cm}
		\caption{Followers distribution.}
		\label{fig:baseline_hatebase_followers}
	\end{subfigure}
	\begin{subfigure}[b]{0.21\textwidth}
		\includegraphics[width=\textwidth]{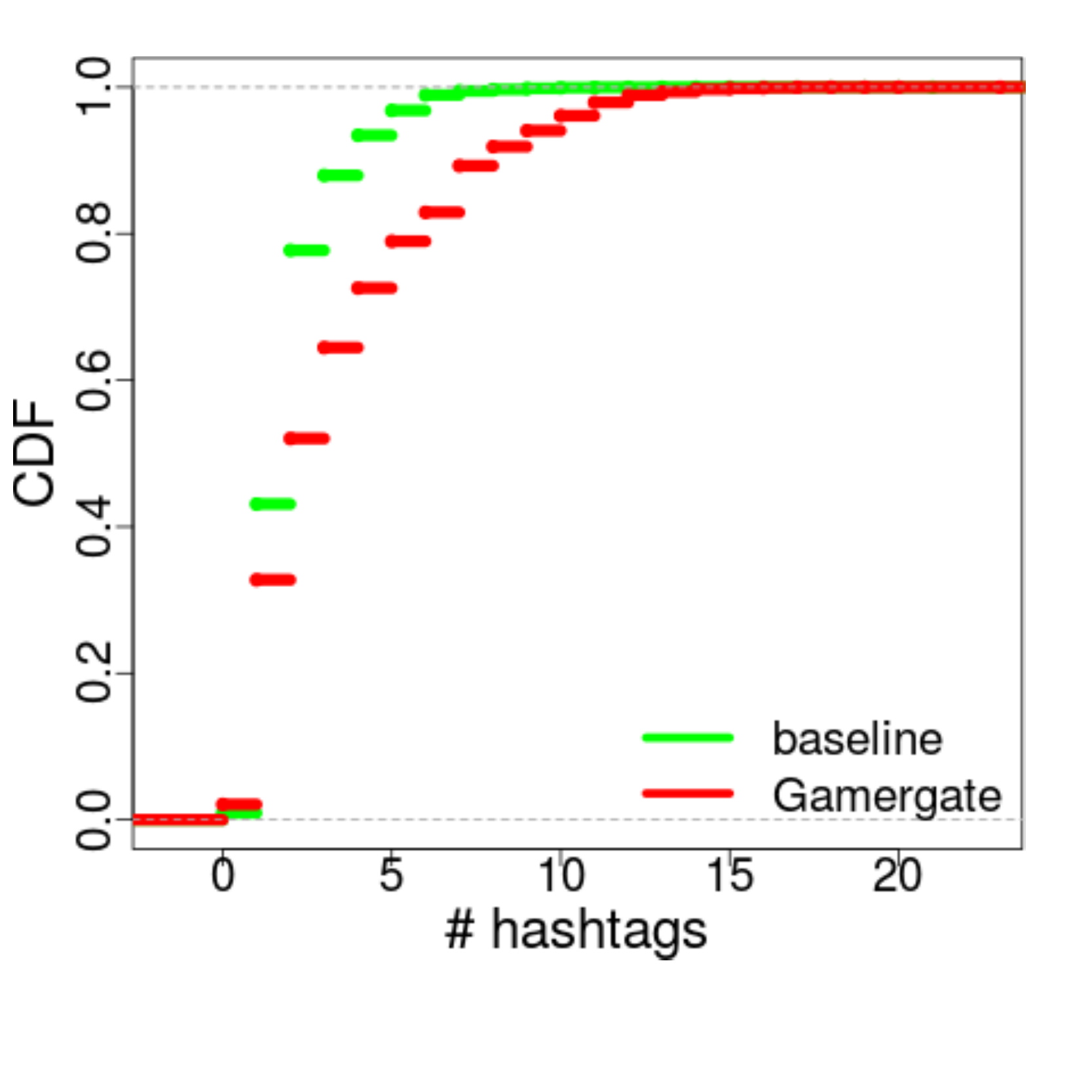}
		\captionsetup{font=scriptsize}
		\vspace{-0.85cm}
		\caption{Hashtags distribution.}
		\label{fig:baseline_hatebase_hashtags}
	\end{subfigure}
	\begin{subfigure}[b]{0.21\textwidth}
		\includegraphics[width=\textwidth]{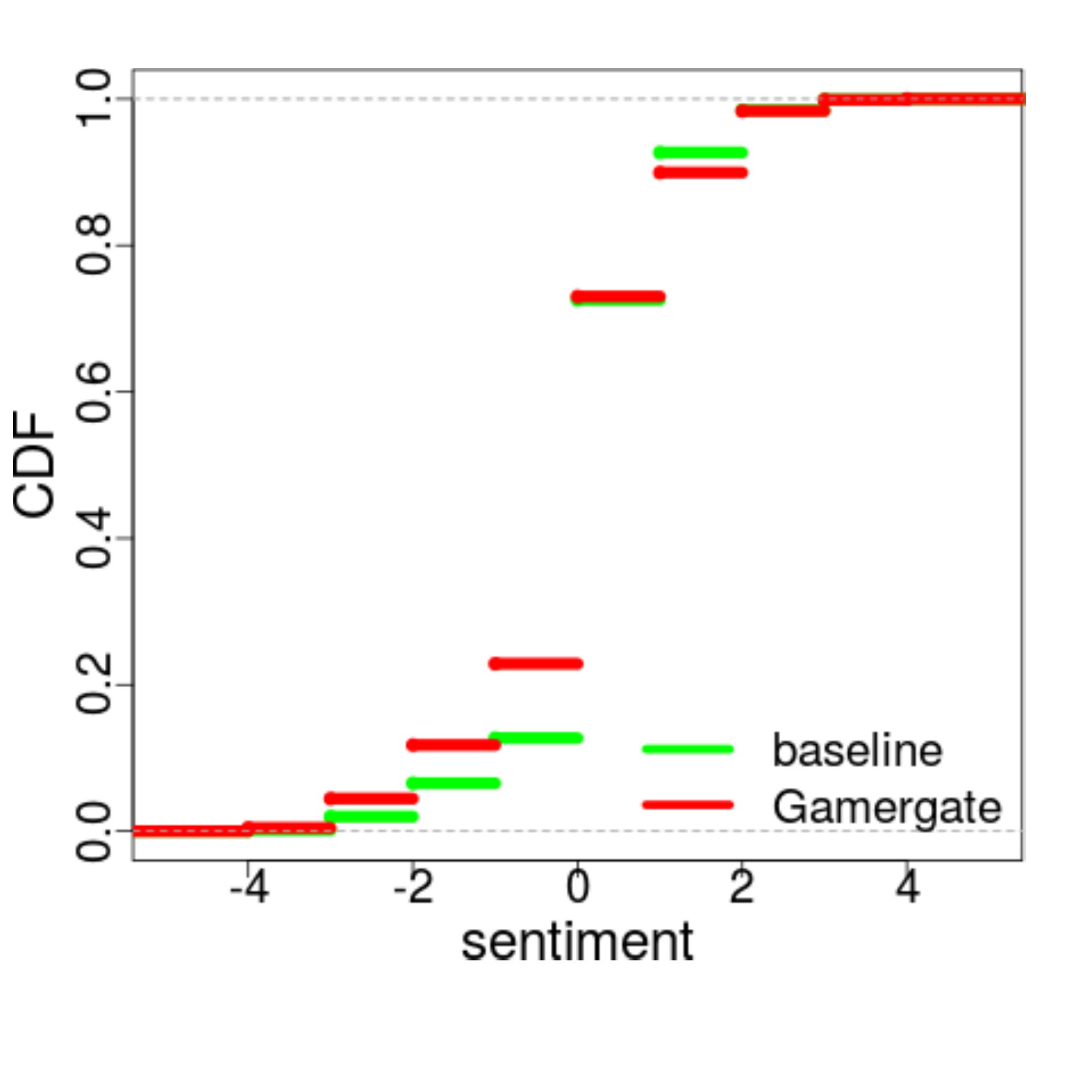}
		\captionsetup{font=scriptsize}
		\vspace{-0.85cm}
		\caption{Sentiment distribution.}
		\label{fig:baseline_hatebase_sentiment}
	\end{subfigure}
	\begin{subfigure}[b]{0.21\textwidth}
		\includegraphics[width=\textwidth]{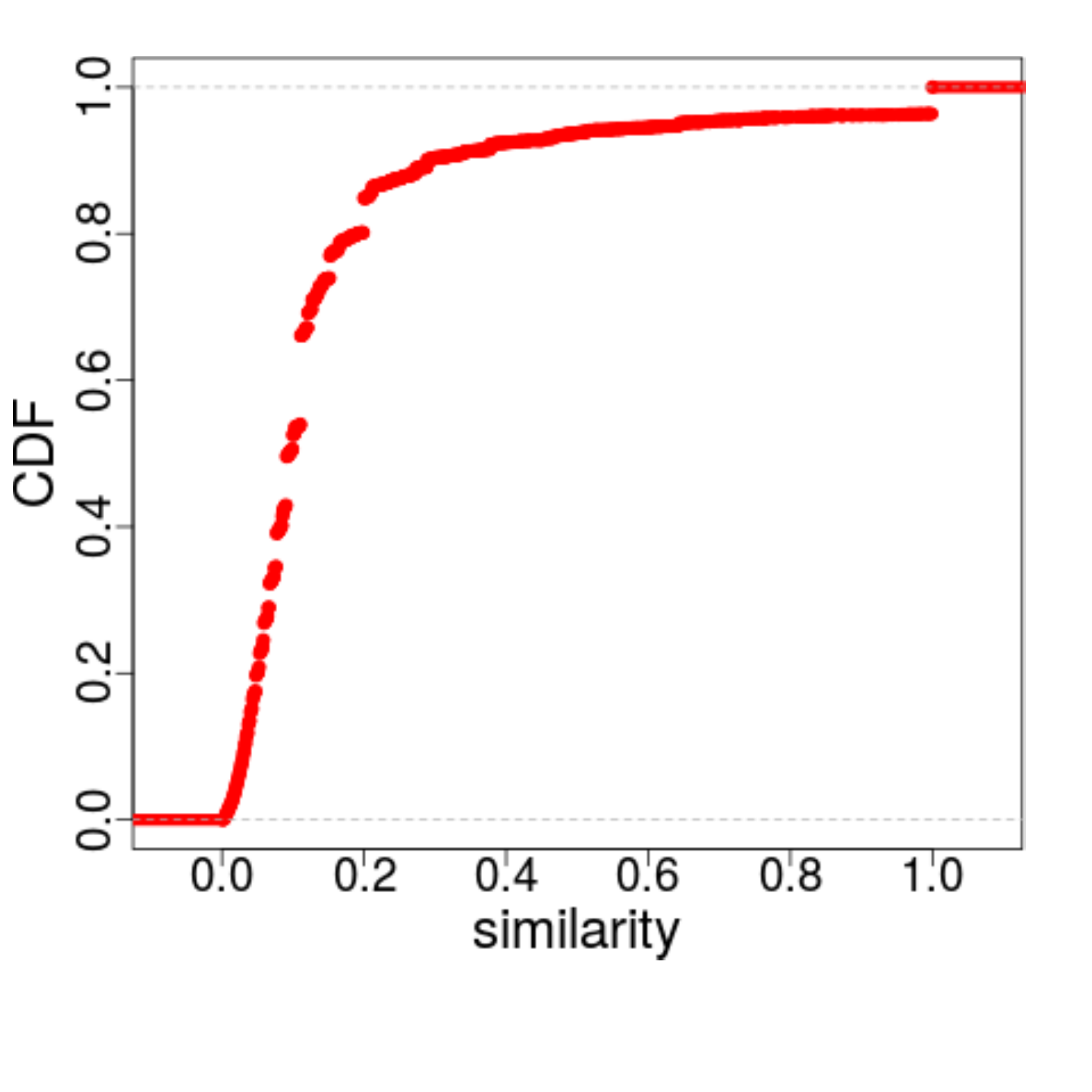}
		\captionsetup{font=scriptsize}
		\vspace{-0.85cm}
		\caption{Similarity distribution.}
		\label{fig:duplications}
	\end{subfigure}
	\vspace{-0.2cm}
	\caption{CDF of (a) Followers, (b) Hashtags, (c) Avg. sentiment, (d) Portion of user's posts with similarity above 0.8.}
	\vspace{-0.3cm}
\end{figure}

\subsection{Data Collection}
Data was collected between June and August 2016, gathering two sets of tweets: (i) a {\bf baseline} of $1$M random tweets, (ii) a {\bf hate-related}, i.e., a set of $650$k tweets collected from the Twitter Streaming API using $309$ hashtags related to bullying and hateful speech.

More specifically, we build the list of $309$ hashtags as follows:
we obtain a $1\%$ sample of all public tweets in the given time window and select all tweets containing \#{GamerGate}.
The Gamergate controversy~\cite{Massanari09102015} is one of the most well documented and mature, large-scale instances of bullying/aggressive behavior that we are aware of.
It originated from alleged improprieties in video game journalism, quickly growing into a larger campaign centered around sexism and social justice~\cite{chatzakou2017measuring}.
With individuals on both sides of the controversy using it, and extreme cases of cyberbullying and aggressive behavior associated with it (e.g., direct threats of rape and murder), \#{GamerGate} serves as a relatively unambiguous hashtag associated with tweets that are likely to involve the type of behavior we are interested in.
We use \#{GamerGate} as a seed for a sort of snowball sampling of other hashtags likely associated with bullying and aggressive behavior; we also include tweets that have one of the $308$ hashtags that appeared in the same tweet as \#{GamerGate}.
Indeed, when manually examining these hashtags, we see that they contain a number of hateful words or hashtags, e.g., \#{IStandWithHateSpeech}, and \#{KillAllNiggers}.

Apart from the hate-related set, we also crawl a random set of tweets to serve as a baseline, as it is less prone to contain any kind of offensive behaviors.
As noted in our previous works~\cite{chatzakou2017measuring, chatzakou2017hypertext}, there are significant differences among the two sets.
To highlight such social and posting activity differences, we show the number of followers, the hashtag usage, and the expressed sentiment.
We observe that users from the hate-related set have more followers compared to the baseline set.
This could be because users with aggressive behavior tend to accumulate more popularity in their network (Fig.~\ref{fig:baseline_hatebase_followers}).
Also, baseline users tweet with fewer hashtags than users from the hate-related dataset (Fig.~\ref{fig:baseline_hatebase_hashtags}), perhaps as the latter use Twitter as a rebroadcasting mechanism aiming at attracting attention to the topic.
Finally, the hate-related tweets contain more negative sentiment since they are more offensive posts (Fig.~\ref{fig:baseline_hatebase_sentiment}).

As expected, the hate-related dataset also contains activity of users who may not be aggressive or hateful, and in fact, may be driving metrics such as popularity in skewed ways.
Therefore, to understand nuanced differences between aggressive or hateful users with normal users, we investigate in more detail behavioral patterns observed in this dataset in Section~\ref{sec:features}. 
More specifically, we analyze user behavior and profile characteristics based on labels (normal, bully, or aggressive) provided by human annotators.

\subsection{Preprocessing}\label{subsec:preprocessing}

We perform several steps to prepare the data for labeling and build ground truth.

\descr{Cleaning.} The first step is to clean the data of noise, i.e., removing numbers, stop words, and punctuations, as well as converting all characters to lower case.

\descr{Removing Spammers.} Previous work has shown that Twitter contains a non-negligible amount of spammers~\cite{Chen2015SpamTweets}, and proposed a number of spam detection tools~\cite{GiatsoglouCSFV15,Wang2010SpamDetectionTwitter}.
Here, we perform a first-level detection of spammers and remove them from our dataset.
We use Wang et al.'s approach~\cite{Wang2010SpamDetectionTwitter}, relying on two main indicators of spam:
(i)~using a large number of hashtags in tweets (to boost visibility), and
(ii)~posting a large number of tweets highly similar to each other.
To find optimal cutoffs for these heuristics, we study both the distribution of hashtags and duplication of tweets. 
For hashtags, we notice that the average number of hashtags within a user's posts ranges in $0$ to $17$.
We experiment with different cutoffs and after a textual inspection on a sample of spam posts, we set the limit to $5$ hashtags, i.e., users with more than $5$ hashtags (on average) per tweet are removed.
Next, we estimate the similarity of a user's tweets via the Levenshtein distance~\cite{Navarro2001ApproximateStringMatching}, i.e., the minimum number of single-character edits needed to convert one string into another, averaging it out over all pairs of the user's tweets.
For a user's $x$ tweets, we calculate the average intra-tweet similarity over a set of $n$ similarity scores, where $n = x (x - 1) / 2$.
If the average intra-tweet similarity is above $0.8$, we exclude the user and their posting activity.
Figure~\ref{fig:duplications} shows that about $5\%$ of users have a high percentage of similar posts and are removed from our dataset.

\subsection{Sessionization}
Cyberbullying usually involves {\em repetitive} actions, thus, we aim to study users' tweets {\em over time}.
Inspired by Hosseinmardi et al.~\cite{Hosseinmardi2015} -- who consider a lower threshold of comments for media sessions extracted from Instagram to be presented in the annotation process -- 
we create, for each user, sets of time-sorted tweets (sessions) by grouping tweets posted close to each other in time.

First, we remove users who are not significantly active, i.e., tweeting less than five times in the 3-month period. 
Then, we use a session-based model where, for each session $S_{i}$, the inter-arrival time between tweets does not exceed a predefined time threshold $t_{l}$. 
We experiment with various values of $t_{l}$ to find an optimal session duration and arrive at a threshold
of 8 hours.
The minimum, median, and maximum length of the resulting sessions (in terms of the number of their included tweets) for the hate-related (i.e., Gamergate) dataset are, respectively, $12$, $22$, and $2.6$k tweets. 
For the baseline set of tweets, they are $5$, $44$, and $1.6$k tweets.

\begin{figure}[!t]
	\centering
	\includegraphics[width=0.43\textwidth]{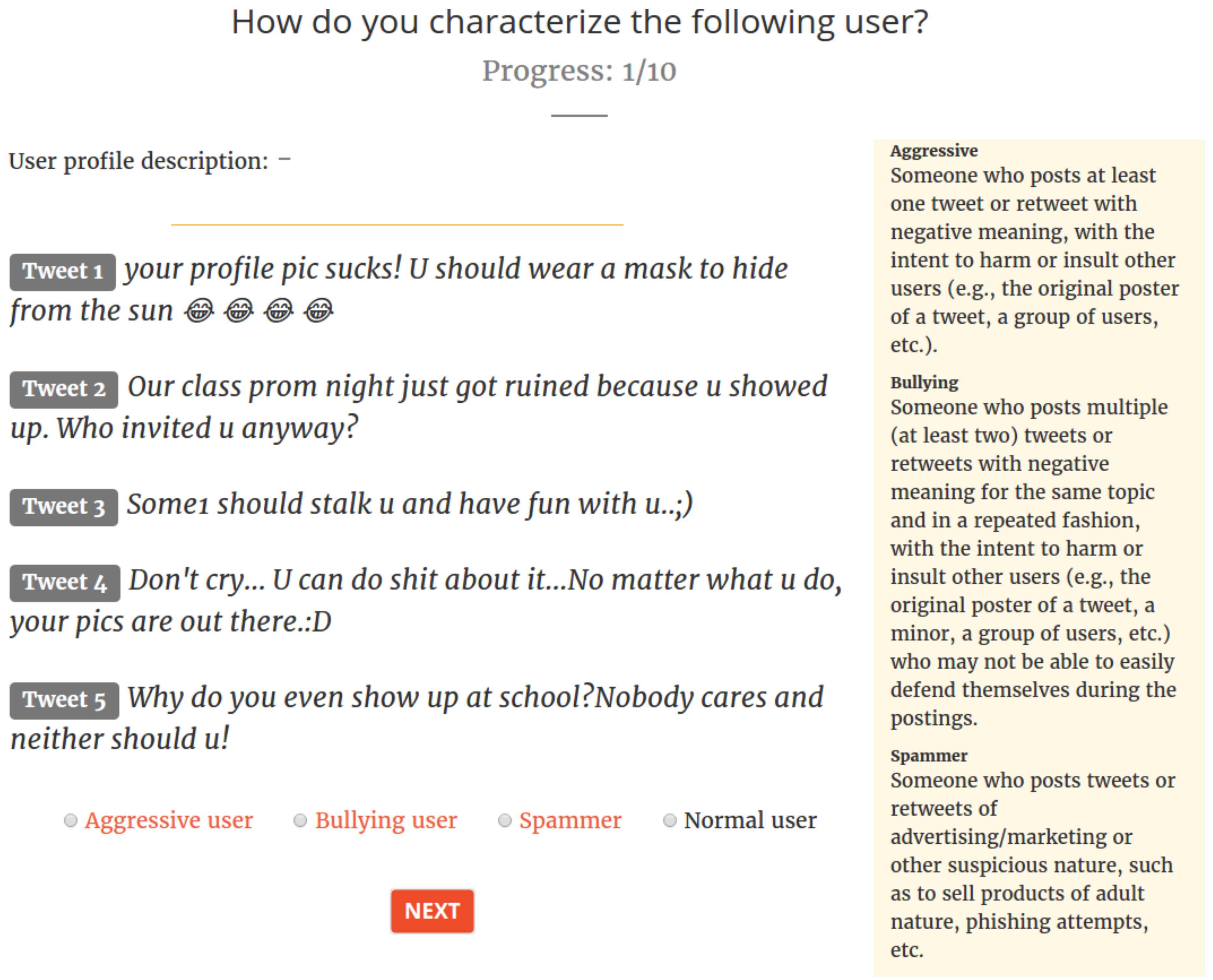}
	\vspace{-0.2cm}
	\caption{Example of the crowdsourcing user interface.}
	\label{fig:crowdsourcing_tool}
	\vspace{-0.3cm}
\end{figure}

Next, we divide sessions in batches, as otherwise they would contain too much information to be carefully examined by a crowdworker within a reasonable period of time.
To find the optimal size of a batch, i.e., the number of tweets per batch, we performed preliminary labeling runs on CrowdFlower, involving $100$ workers each, using batches of exactly $5$, $5$-$10$, and $5$-$20$ tweets. 
Our intuition is that increasing the batch size provides more context to the worker to assess if a poster is acting in an aggressive or bullying behavior, however, too many tweets might confuse them.
The best results with respect to labeling agreement -- i.e., the number of workers that provide the same label for a batch -- occur with $5$-$10$ tweets per batch.
Therefore, we eliminate sessions with fewer than $5$ tweets, and further split those with more than $10$ tweets (preserving the chronological ordering of their posted time).
In the end, we arrive at $1,500$ batches.
We also note that we maintain the same number of batches for both the hate-related and baseline tweets.

\subsection{Crowdsourced Labeling}

We now present the design of our crowdsourcing labeling process, performed on CrowdFlower. 

\descr{Labeling.} Our goal is to label each Twitter user -- {\em not} single tweets -- as \textit{normal}, \textit{aggressive}, \textit{bullying}, or \textit{spammer} by analyzing their batch(es) of tweets.
Note that we also allow for the possibility that a user is spamming and has passed our basic spam filtering.
Based on previous research~\cite{Smith2008CyberbullyingNature,Tokunaga2010CriticalReviewCyberbullyingVictimization,grigg2010cyber}, workers are provided with the following definitions of aggressive, bullying, and spam behaviors:

\begin{squishlist}
\item \emph{aggressor:} someone who posts at least one tweet or retweet with negative meaning, with the intent to harm or insult other users (e.g., the original poster of a tweet, a group of users, etc.);
\item \emph{bully:} someone who posts multiple tweets or retweets ($\ge$$2$) with negative meaning for the same topic and in a repeated fashion, with the intent to harm or insult other users (e.g., the original poster of a tweet, a minor, a group of users, etc.) who may not be able to easily defend themselves during the postings;
\item \emph{spammer:} someone who posts texts of advertising / marketing or other suspicious nature, such as to sell products of adult nature, and phishing attempts. 
\end{squishlist}

\descr{CrowdFlower Task.}
We redirect employed crowd workers to an online survey tool we developed.
First, they are asked to provide basic demographic information: gender, age, nationality, education level, and annual income.
In total, $30\%$ are female and $70\%$ male, while their claimed educational level is spread between secondary education ($18.4\%$), bachelor degree ($35.2\%$), master ($44\%$), and PhD ($2.4\%$).
One third ($35.5\%$) claim an income level below \euro10k, $\sim$$20\%$ between \euro10k-\euro20k and the rest between \euro20k-\euro100k.
Regarding age, $27\%$ are 18-24, $30\%$ are 25-31, $21\%$ are 32-38, $12\%$ are 39-45, and the remainder above 45 years old.
They come from 56 different countries, with significant participation of users from USA, Venezuela, Russia, and Nigeria.
Overall, the annotators from the top 10 countries contribute $75\%$ of all annotations.
 
We then ask workers to label $10$ batches, one of which is a control case (details below).
We also provide them with the user profile description (if any) of the Twitter user they are labeling and the definition of aggressive, bullying, and spammer behaviors.
Figure~\ref{fig:crowdsourcing_tool} presents an example of the interface.
The workers rated the instructions given to them, as well as the overall task, as very good with an overall score of 4 out 5.

\descr{Results.}
Overall, we recruit 834 workers. 
They are allowed to participate only once to eliminate behavioral bias across tasks and discourage rushed tasks.
Each batch is labeled by $5$ different workers, and, similar to~\cite{Hosseinmardi2015},~\cite{Nobata2016AbusiveLanguageDetection}, a majority vote is used to decide the final label.
We receive $1,303$ annotated batches, comprising $9,484$ tweets in total.
$4.5\%$ of users are labeled as bullies, $3.4\%$ as aggressors, $31.8\%$ as spammers, and $60.3\%$ as normal.
Overall, abusive users (i.e., bullies and aggressors) make up about 8\% of our dataset, which mirrors observations from previous studies (e.g., in~\cite{kayes2015ya-abuse} $9\%$ of the users in the examined dataset exhibits bad behavior, while in~\cite{Blackburn2012Cheaters} $7\%$ of users cheated).
Thus, we believe our ground truth dataset contains a representative sample of aggressive/abusive content.

\descr{Annotator reliability.}
To assess the reliability of our workers, we use (i)~the inter-rater reliability measure, and (ii)~control cases.
We find the inter-rater agreement to be $0.54$.
We also use control cases to ensure worker ``quality'' by manually annotating three batches of tweets.
During the annotation process, each worker is given a set of batches to annotate, one of which is a randomly selected control case: the annotation of these control cases is used to assess their ability to adequately annotate for the given task.
We find $66.5\%$ accuracy overall (i.e., the percent of correctly annotated control cases).
More specifically, $83.75\%$ accuracy for spam, $53.56\%$ for bully, and $61.31\%$ for aggressive control cases.

\section{Feature Extraction}\label{sec:features}

\begin{table}[!t]
\begin{center}
\scalebox{0.72}{
\begin{tabular}{|l|l|}
\hline
{\bf Type}				&	{\bf Feature}											\\
\hline
User				&	avg. \# posts, \# days since account creation, verified account\\
(total: 10)
			    	&	\# subscribed lists, posts' interarrival time, default profile image?\\
			    	&	statistics on sessions: total number, avg., median, and STD. of their size\\
\hline
Textual			&	avg. \# hashtags, avg. \# emoticons, avg. \# upper cases, \# URLs\\
(total: 9)				&	avg. sentiment score, avg. emotional scores, hate score\\
				&   avg. word embedding score, avg. curse score	\\
\hline
Network			&	\# friends, \# followers, hubs,  (d=\#followers/\#friends), authority\\
(total: 11)				&	avg. power diff. with mentioned users,  clustering coefficient, reciprocity \\
				&	eigenvector centrality, closeness centrality, louvain modularity\\
\hline
\end{tabular}
}
\end{center}
\caption{Features considered in the study.}
\label{tab:features}
\vspace{-0.5cm}
\end{table}%

In this section, we focus on user-, text-, and network-based features that can be subsequently used in the modeling of user behaviors identified in the dataset.
Next, we detail the features from each category, summarized in Table~\ref{tab:features}.
To examine the significance of differences among the distributions presented next, we use the two-sample Kolmogorov-Smirnov test, a non-parametric statistical test, to compare the probability distributions of different samples.
We consider as statistically significant all cases with $p < 0.05$.

\begin{figure}[!t]
	\centering
	\begin{subfigure}[b]{0.21\textwidth}
		\captionsetup{font=scriptsize}
		\includegraphics[width=\textwidth]{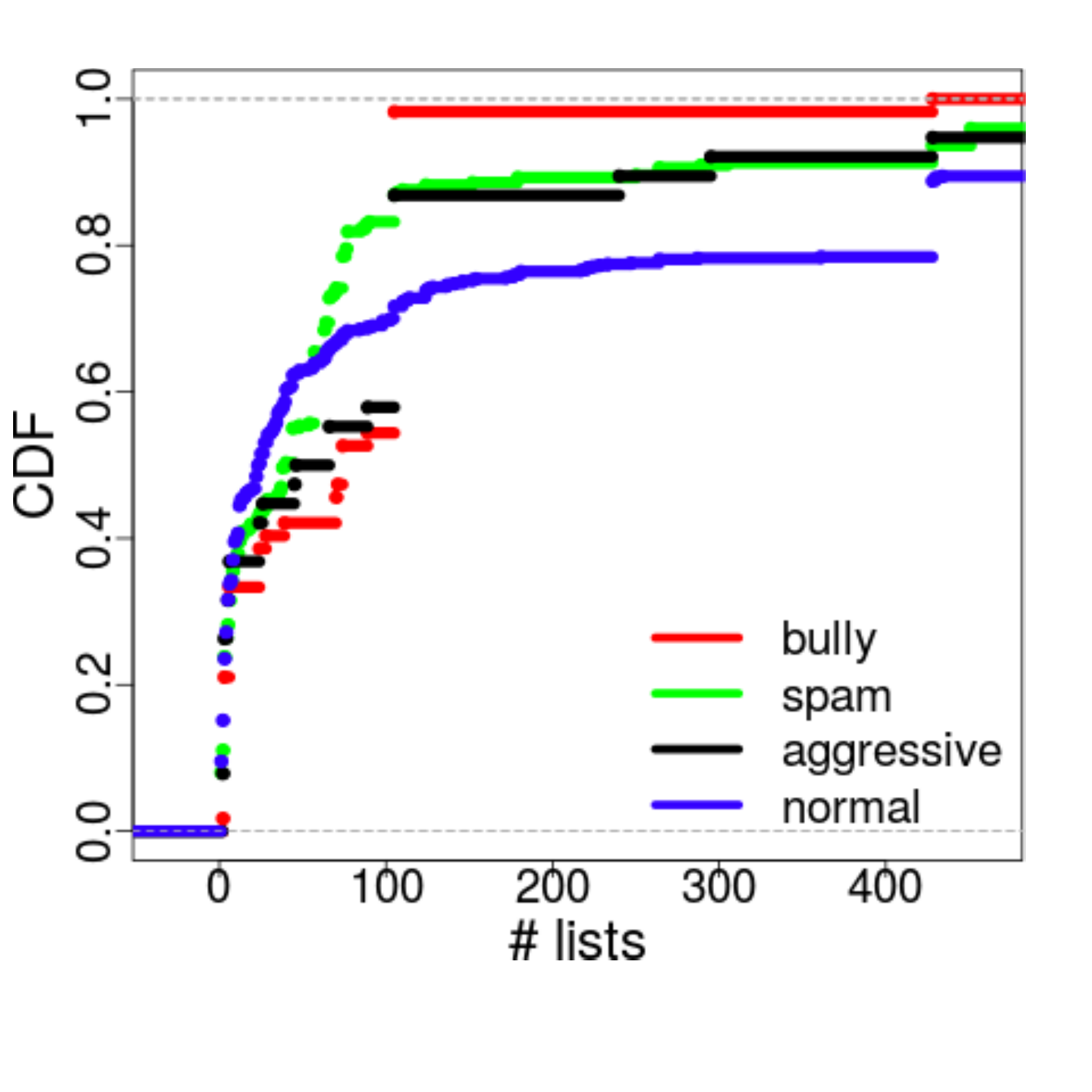}
		\vspace{-0.85cm}
		\caption{Lists distribution.}
		\label{fig:ecdf-cntlists}
	\end{subfigure}
	\begin{subfigure}[b]{0.21\textwidth}
		\captionsetup{font=scriptsize}
		\includegraphics[width=\textwidth]{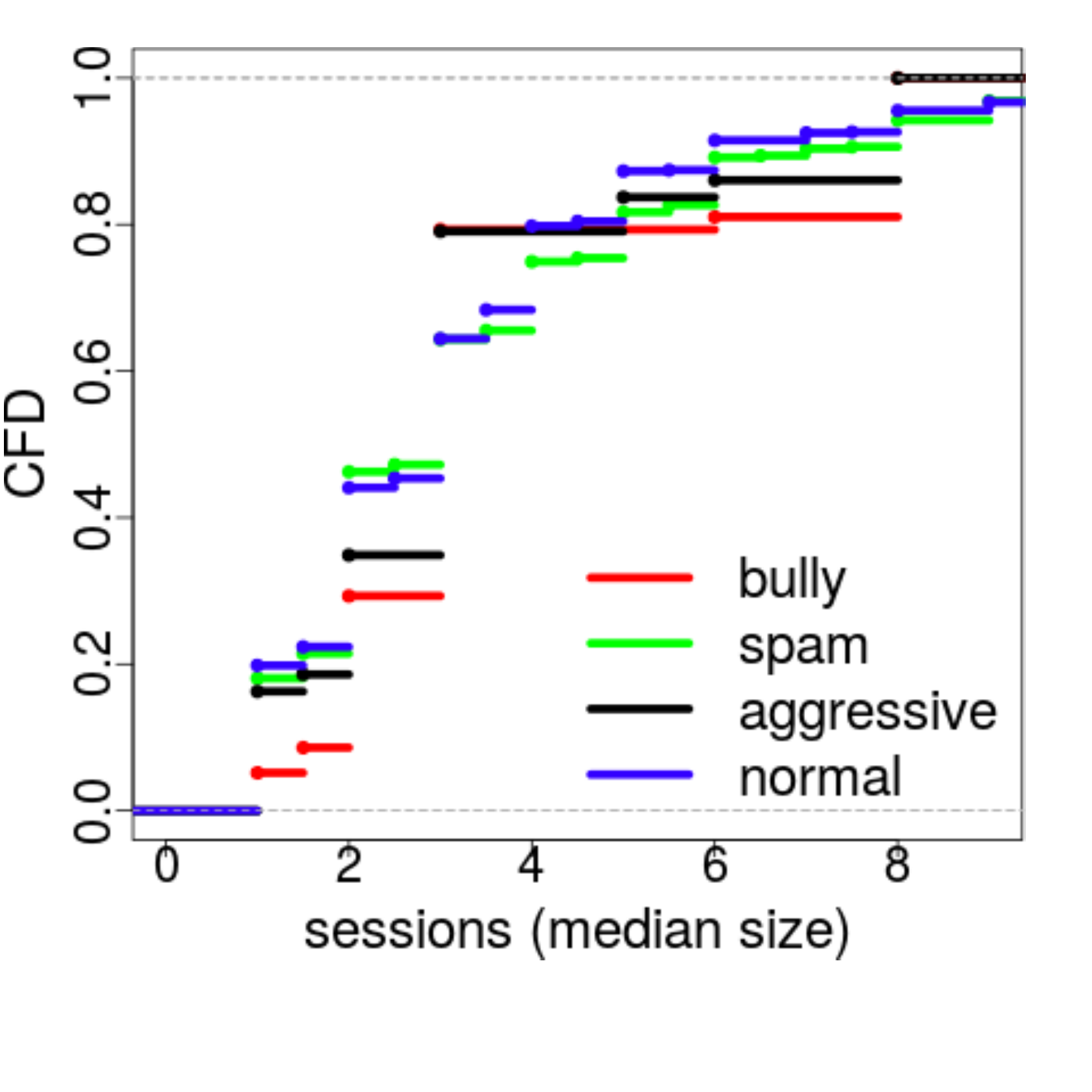}
		\vspace{-0.85cm}
		\caption{Sessions distribution.}
		\label{fig:ecdf-sessions_median}
	\end{subfigure}	
	\begin{subfigure}[b]{0.21\textwidth}
		\captionsetup{font=scriptsize}
		\includegraphics[width=\textwidth]{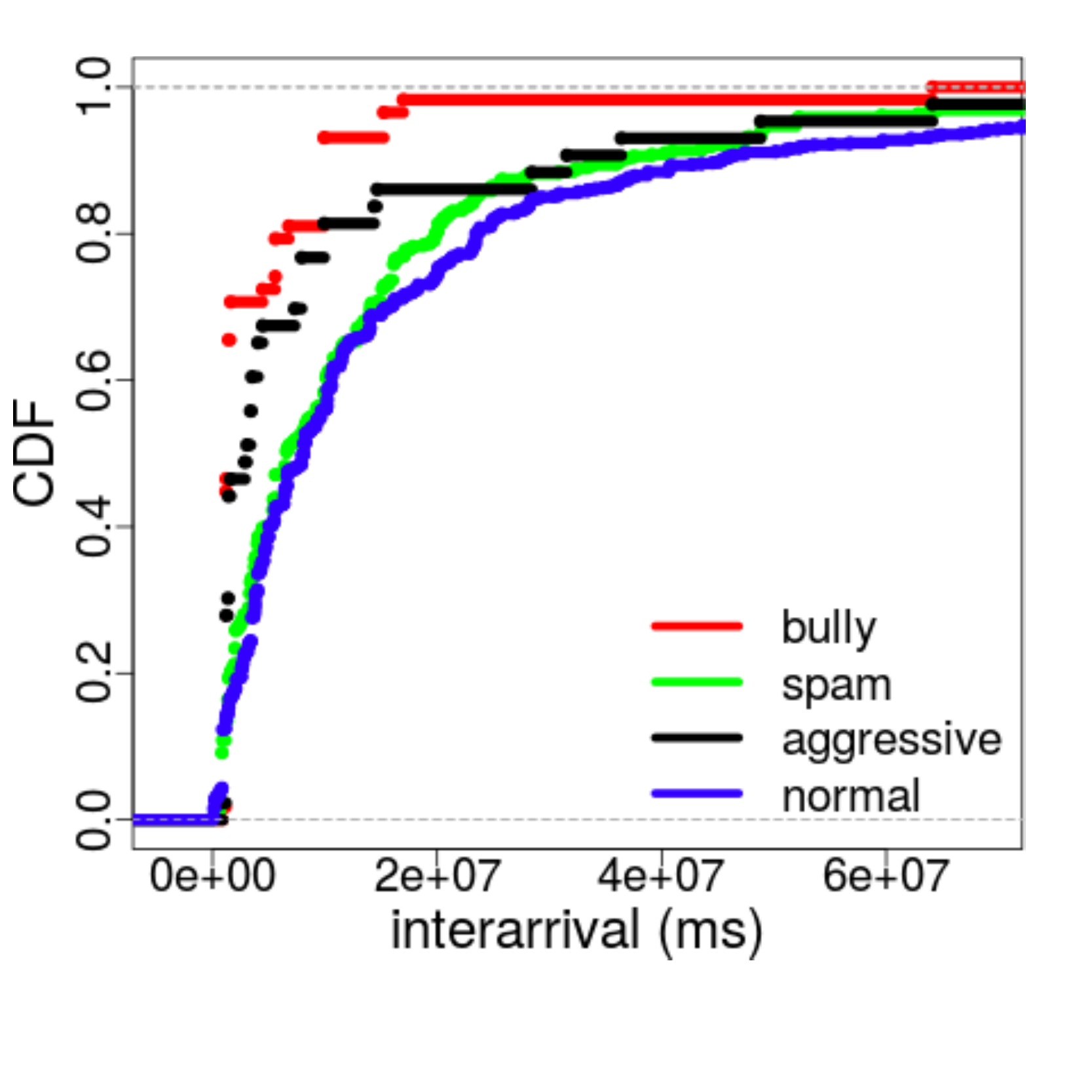}
		\vspace{-0.85cm}
		\caption{Interarrival time distribution.}
		\label{fig:ecdf-interarrival-all}
	\end{subfigure}	
	\begin{subfigure}[b]{0.21\textwidth}
		\captionsetup{font=scriptsize}
		\includegraphics[width=\textwidth]{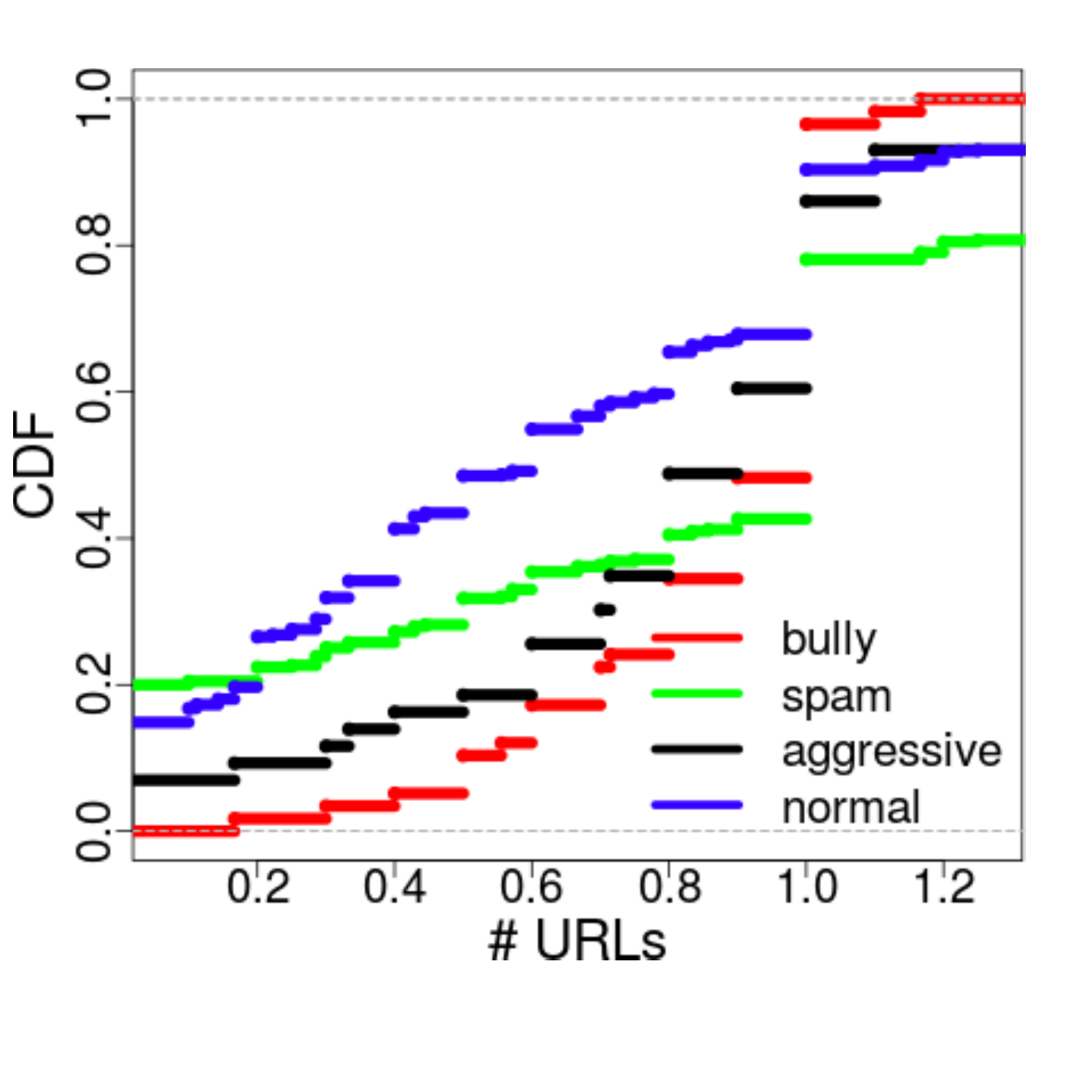}
		\vspace{-0.85cm}
		\caption{URLs distribution.}
		\label{fig:ecdf-cntUrls}
	\end{subfigure}
	\begin{subfigure}[b]{0.21\textwidth}
		\captionsetup{font=scriptsize}
		\includegraphics[width=\textwidth]{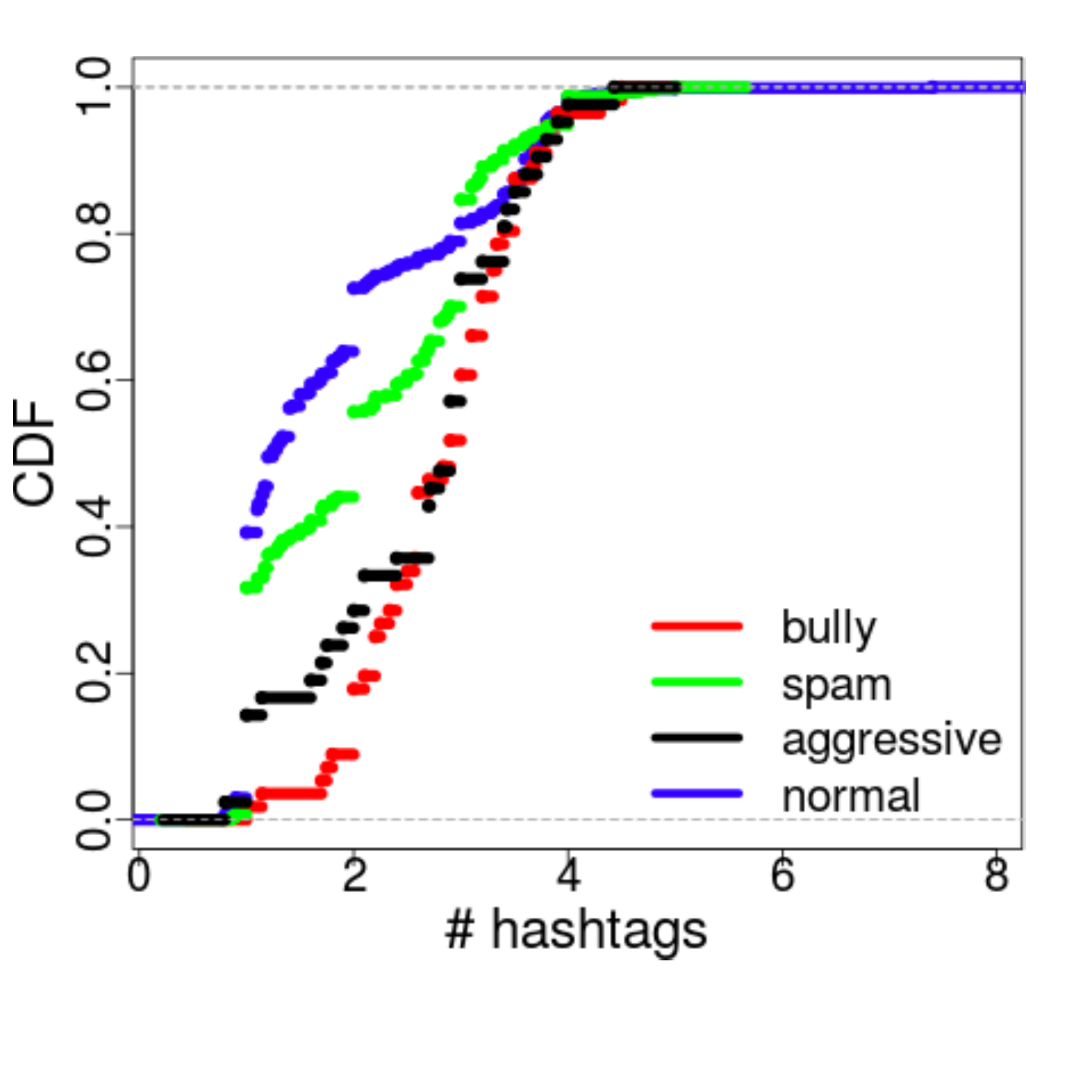}
		\vspace{-0.85cm}
		\caption{Hashtags distribution.}
		\label{fig:ecdf-cntHashtags}
	\end{subfigure}
	\begin{subfigure}[b]{0.21\textwidth}
		\captionsetup{font=scriptsize}
		\includegraphics[width=\textwidth]{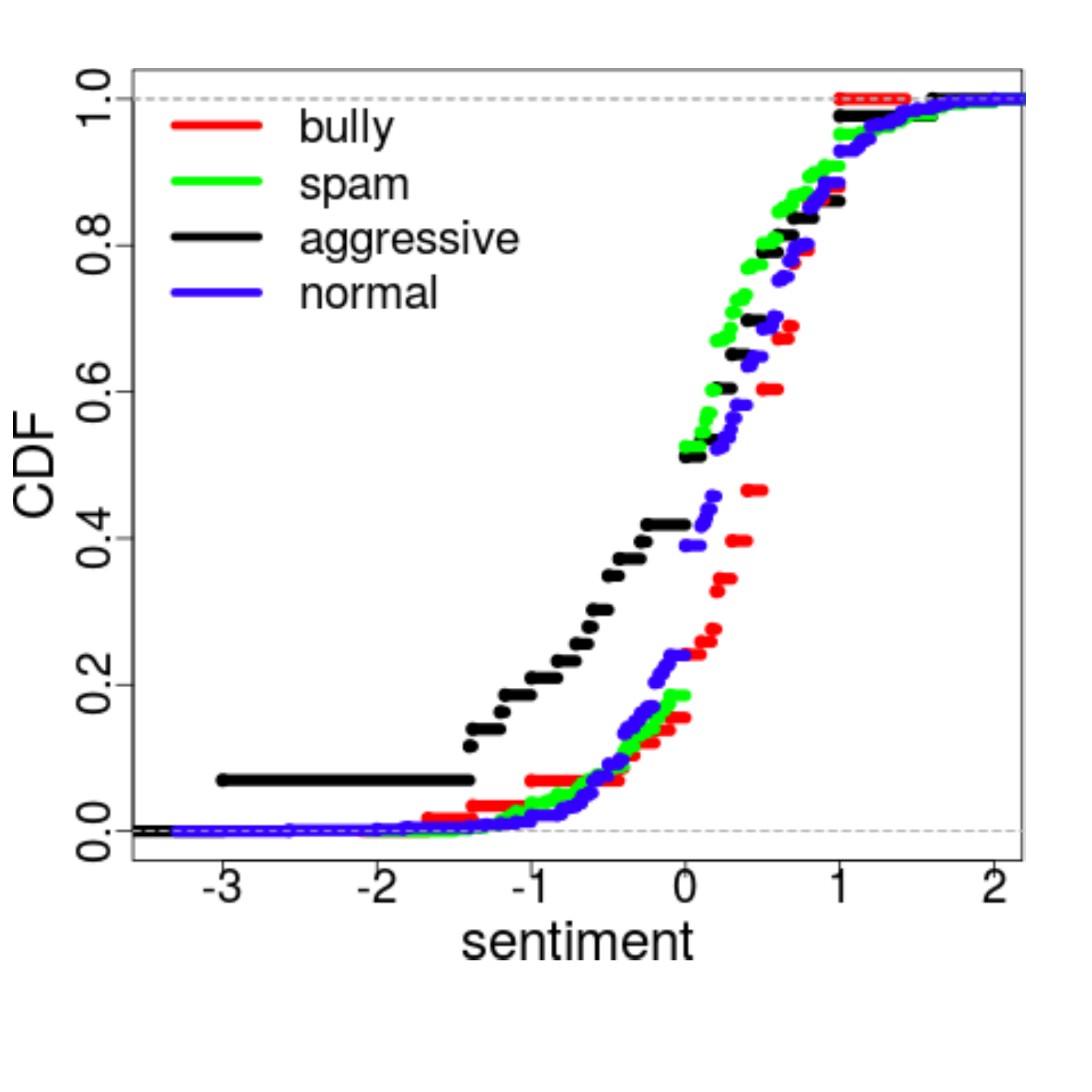}
		\vspace{-0.85cm}
		\caption{Sentiment distribution.}
		\label{fig:ecdf-sentiment}
	\end{subfigure}
	\vspace{-0.2cm}
	\caption{(a) Lists, (b) Sessions size, (c) Interarrival time, (d) URLs, (e) Hashtags, and (f) Sentiment distribution.}
	\vspace{-0.4cm}
\end{figure}

\subsection{User-based features}

\descr{Basics.} We experiment with various user-based features; i.e., features extracted from a user's profile.
Features in this category include the number of tweets a user has made, the age of his account (i.e., number of days since its creation), the number of lists subscribed to, if the account is verified or not (i.e., acknowledged by Twitter as an account linked to a user of ``public interest''), and whether or not the user still uses the default profile image.
A representative example is shown in Figure~\ref{fig:ecdf-cntlists}, which plots the CDF of the number of subscribed lists for each of the four behaviors we examine (the maximum number of lists is $4,327$, but we trim the plot at $500$ for readability).
The median (max) number of lists for bullying, spam, aggressive, and normal users is $24$ ($428$), $57$ ($3,723$), $40$, ($1,075$), and $74$ ($4,327$), respectively.
We note the difference in the participation of groups from each class of users, with normal users signing up to more lists than the other types of users.

\descr{Session statistics.} Here, we consider the number of sessions produced by a user from June to August and we estimate the average, median and standard deviation of the size of each users' sessions.
Figure~\ref{fig:ecdf-sessions_median} shows the CDF of the median number of sessions for the $4$ behavior classes.
Comparing the distributions among the bullies and aggressors to the normal users, we observe that the differences are not statistically significant with $D$$=$$0.16052$ and $D$$=$$0.14648$ for bully vs. normal, and aggressors vs. normal, respectively.

\descr{Inter-arrival time.} Figure~\ref{fig:ecdf-interarrival-all} plots the CDF of users posts' inter-arrival time.
We observe that bullies and aggressors tend to have less waiting time in their posting activity compared to the spam and normal users, which is in alignment with results in~\cite{Hosseinmardi2015}.

\subsection{Text-based features}

For text-based features, we look deeper into a user's tweeting activity by analyzing specific attributes that exist in his tweets.

\descr{Basics.}
We consider some basic metrics across a user's tweets: the number of hashtags used, uppercase text (which can be indicative of intense emotional state or `shouting'), number of emoticons, and URLs.
For each of these, we take the average over all tweets in a users' annotated batch.
Figure~\ref{fig:ecdf-cntUrls} depicts the CDF of the average number of URLs for the different classes of users.
The median value for bully, spam, aggressive, and normal users is $1$, $1$, $0.9$, and $0.6$, respectively.
The maximum number of URLs between users also varies: for bully and aggressive users it is $1.17$ and $2$, respectively, while for spam and normal users it is $2.38$ and $1.38$. 
Thus, normal users tend to post fewer URLs than the other 3 classes.
Also, from Figure~\ref{fig:ecdf-cntHashtags} we observe that aggressive and bully users have a propensity to use more hashtags within their tweets, as they try to disseminate their attacking message to more individuals or groups.

\descr{Word embedding.}
Word embedding allows finding both semantic and syntactic relation of words, which permits the capturing of more refined attributes and contextual cues that are inherent in human language.
E.g., people often use irony to express their aggressiveness or repulsion.
Therefore, we use Word2Vec~\cite{MikolovWordEmbedding2013}, an unsupervised word embedding-based approach to detect semantic and syntactic word relations.
Word2Vec is a two-layer neural network that operates on a set of texts to: 1)~initially establish a vocabulary based on the words included in such set more times than a user-defined threshold (to eliminate noise), 2)~apply a learning model to input texts to learn the words' vector representations in a $D$-dimensional, user-defined space, and 3)~output a vector representation for each word encountered in the input texts.
Based on~\cite{MikolovWordEmbedding2013} $50$-$300$ dimensions can model hundreds of millions of words with high accuracy.
Possible methods to build the actual model are: 1)~CBOW (i.e., Continuous bag of words), which uses context to predict a target word, and 2)~Skip-gram, which uses a word to predict a target context.
Skip-gram works well with small amounts of training data and handles rare words or phrases well, while CBOW shows better accuracy for frequent words and is faster to train.

Here, we use Word2Vec to generate features to better capture the context of the data at hand.
We use a pre-trained model with a large scale thematic coverage (with 300 dimensions) and apply the CBOW model due to its better performance regarding the training execution time.
Finally, having at hand the vector representations of all input texts' words, the overall vector representation of an input text is derived by averaging all the vectors of all its comprising words.
Comparing the bully distribution with the normal one we conclude to $D$$=$$0.094269$ and $p$$=$$0.7231$, while in the aggressive vs. normal distribution comparison $D$$=$$0.11046$ and $p$$=$$0.7024$, thus in both cases the differences are not statistically significant.
 
\descr{Sentiment.}
Sentiment has already been considered during the process of detecting abusive behavior in communications among individuals, e.g.,~\cite{Nahar2012SentimentAnalysisDetectionCyberBullying}.
To detect sentiment, we use the SentiStrength tool~\cite{sentistrength}, which estimates the positive and negative sentiment (on a [-4, 4] scale) in short texts, even for informal language often used on Twitter.
First, however, we evaluate its performance by applying it on an already annotated dataset with $7,086$ tweets~\cite{sentimentAnnotatedDataset}.
The overall accuracy is 92\%, attesting to its efficacy for our purposes.
Figure~\ref{fig:ecdf-sentiment} plots the CDF of average sentiment for the 4 user classes.
Even though we observe a distinction among the aggressive and the rest of classes, this is not the case when comparing the remaining three classes, where similar behavior is observed concerning the expressed sentiments.
More specifically, comparing the distributions of the aggressive class with the normal, and bully with normal, they are statistically different ($D$$=$$0.27425$ and $D$$=$$0.27425$, respectively).
We also attempt to detect more concrete emotions, i.e., anger, disgust, fear, joy, sadness, and surprise based on the approach presented in~\cite{Chatzakou2013EmotionallyDrivenClustering}.
Comparing the distributions of the abusive classes with the normal, in most of cases we observe no statistical difference.
For anger, even though the aggressive and normal distributions are statistically different ($D$$=$$0.21515$), the bully and normal users are not ($D$$=$$0.080029$ and $p$$=$$0.88$).

\descr{Hate and Curse words.}
Additionally, we wanted to specifically examine the existence of hate speech and curse words within tweets.
For this purpose, we use the Hatebase database (HB)~\cite{hatebase}, which is a crowdsourced list of hate words.
Each word in the HB is additionally scored on a [0, 100] scale indicating how hateful the word is.
Finally, a list of swear words~\cite{swearwords} is also used in a binary fashion; i.e., we set a variable to true if a tweet contained any word in the list, and false otherwise.
Even though these lists can be useful in categorizing general text as hateful or aggressive, they are not well suited for classifying tweets as they are short and typically include modified words, URLs and emoticons.
Overall, we find that bully and aggressive users have a minor bias towards using such words, but they are not significantly different from normal users' behavior.

\begin{figure}[!t]
	\centering
	\begin{subfigure}[b]{0.21\textwidth}
			\captionsetup{font=scriptsize}
			\includegraphics[width=\textwidth]{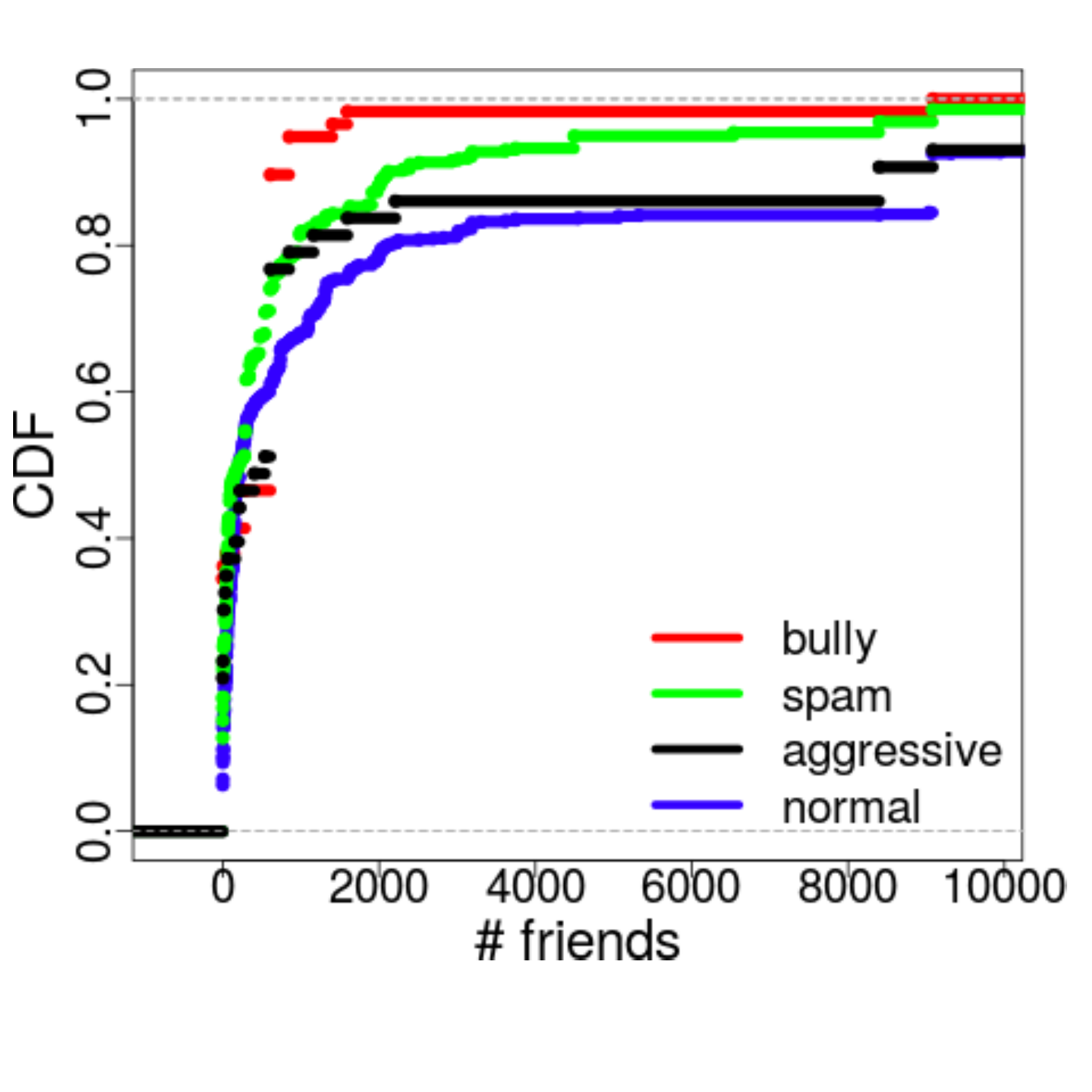}
			\vspace{-0.85cm}
			\caption{Friends distribution.}
			\label{fig:ecdf-friends-features}
	\end{subfigure}
	\begin{subfigure}[b]{0.21\textwidth}
			\captionsetup{font=scriptsize}
			\includegraphics[width=\textwidth]{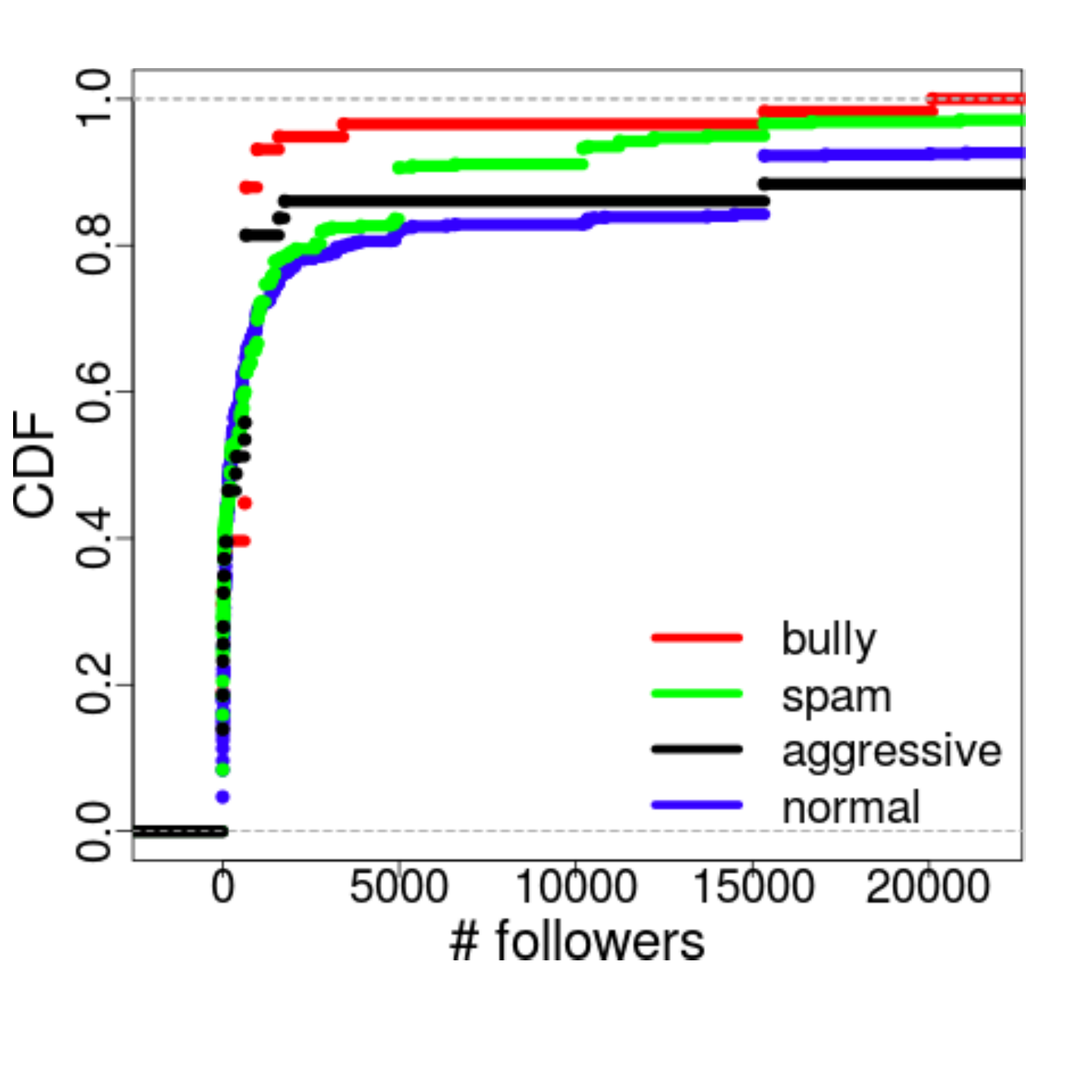}
			\vspace{-0.85cm}
			\caption{Followers distribution.}
			\label{fig:ecdf-followers-features}
	\end{subfigure}
	\begin{subfigure}[b]{0.21\textwidth}
		\captionsetup{font=scriptsize}
		\includegraphics[width=\textwidth]{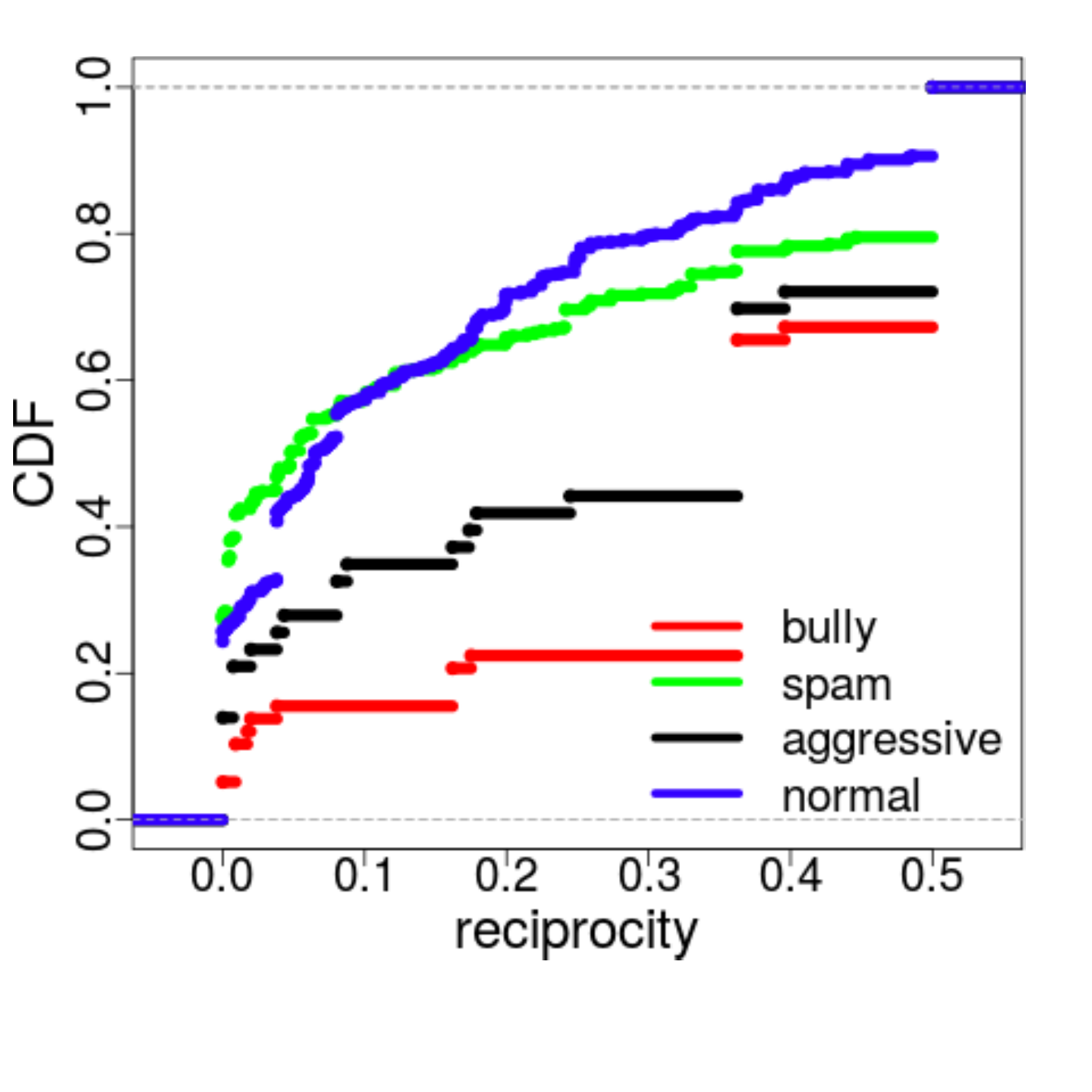}
		\vspace{-0.85cm}
		\caption{Reciprocity distribution.}
		\label{fig:ecdf-reciprocity}
	\end{subfigure}
	\begin{subfigure}[b]{0.21\textwidth}
		\captionsetup{font=scriptsize}
		\includegraphics[width=\textwidth]{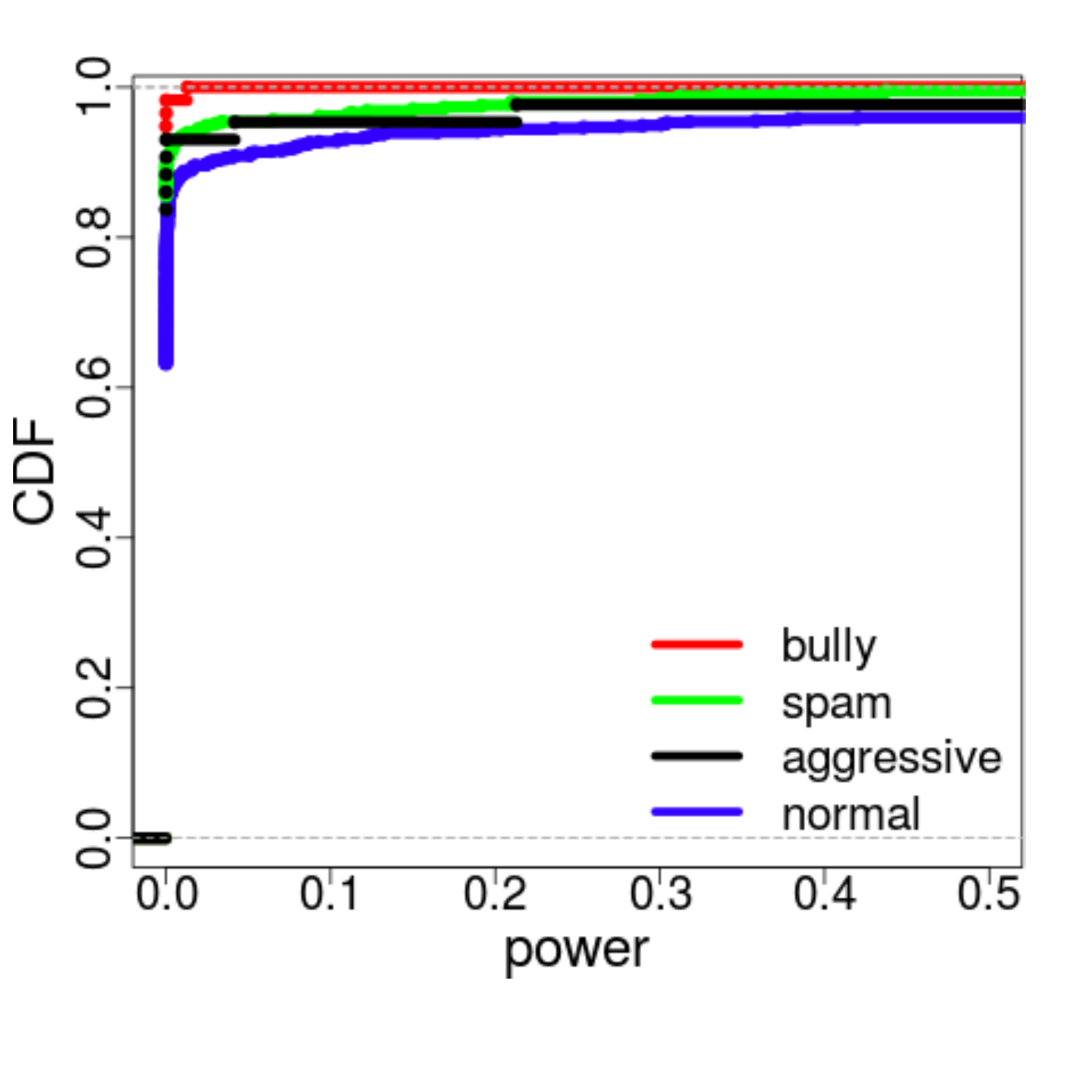}
		\vspace{-0.85cm}
		\caption{Power difference distribution.}
		\label{fig:ecdf-power}
	\end{subfigure}
	\vspace{-0.2cm}
	\caption{(a) Friends, and (b) Followers distribution, (c) Avg. distribution for Reciprocities, and (d) Power difference.}
	\vspace{-0.4cm}
\end{figure}

\subsection{Network-based features}
Twitter social network plays a crucial role in diffusion of useful information and ideas, but also of negative opinions, rumors, and abusive language (e.g.,~\cite{Jin2013EpidemiologicalModelingNewsAndRumors,Pfeffer2014NegativeWordToMouthDynamics}).
We study the association between aggressive or cyberbullying behavior and the position of users in the Twitter network of friends and followers.
The network is comprised of about $1.2M$ users and $1.9M$ friend (i.e., someone who is followed by a user) or follower (i.e., someone who follows a user) edges, with $4.934$ effective diameter, $0.0425$ average clustering coefficient, and $24.95\%$ and $99.99\%$ nodes in the weakest and largest component, respectively.
Users in such a network can have a varying degree of embeddedness with respect to friends or followers, reciprocity of connections, connectivity with different parts of the network, etc.

\descr{Popularity.}
The popularity of a user can be defined in different ways.
For example, the number of friends or followers (out- or in-degree centrality), and the ratio of the two measures (since Twitter allows users to follow anyone without their approval, the ratio of followers to friends can quantify a user's popularity).
These measures quantify the opportunity for a user to have a positive or negative impact in his ego-network in a direct way.
Figures~\ref{fig:ecdf-friends-features} and~\ref{fig:ecdf-followers-features} indicate that bullies have fewer friends and followers than the other user categories, with normal users having the most friends.

\descr{Reciprocity.}
This metric quantifies the extent to which users reciprocate the follower connections they receive from other users.
The average reciprocity in our network is $0.2$.
Figure~\ref{fig:ecdf-reciprocity} shows that the user classes considered have different distributions, with the bully and aggressive users being more similar (i.e., higher number of reciprocities) than the normal or spammers.
Reciprocity as a feature has also been used in~\cite{Hosseinmardi2014TowardsUnderstandingCyberbullying}, but in an interaction-based graph using likes in posts.
Here, we investigate the fundamental reciprocity in Twitter friendship; the first to do this in the context of bullying.

\descr{Power Difference.}
A recent study~\cite{Pieschl2013241} found that the emotional and behavioral state of victims depend on the power of their bullies, e.g., more negative emotional experiences were observed when more popular cyberbullies conducted the attack, and the high power difference with respect to status in the network has been shown to be a significant characteristic of bullies~\cite{Corcoran2015CyberbullyingOrCyberAggression}.
Thus, we consider the power difference between a tweeter and his mentions.
In fact, a further analysis of a user's mentioned users could reveal possible victims or bystanders of his aggressive or bullying behavior.
To this end, we compute the difference in power a user has with respect to the users he mentions in his posts, in terms of their respective followers/friends ratio.
Figure~\ref{fig:ecdf-power} shows the distribution of this power difference (we note that the maximum power difference is 20, but we trim the plot for readability).
The difference in power between the aggressive (bully) and normal users is statistically significant ($D$$=$$0.22133$ and $D$$=$$0.31676$, respectively).

\descr{Centrality Scores.}
We also investigate users' position in their network through more elaborate metrics such as hub, authority, eigenvector and closeness centrality, that measure influence in their immediate and extended neighborhood, as well as connectivity.

\descr{Hubs and Authority.}
A node's hub score is the sum of the authority score of the nodes that point to it, and authority shows how many different hubs a user is connected with~\cite{Kleinberg1999HubsAuthorities}.

\begin{figure}[!t]
	\centering
	\begin{subfigure}[b]{0.21\textwidth}
		\captionsetup{font=scriptsize}
		\includegraphics[width=\textwidth]{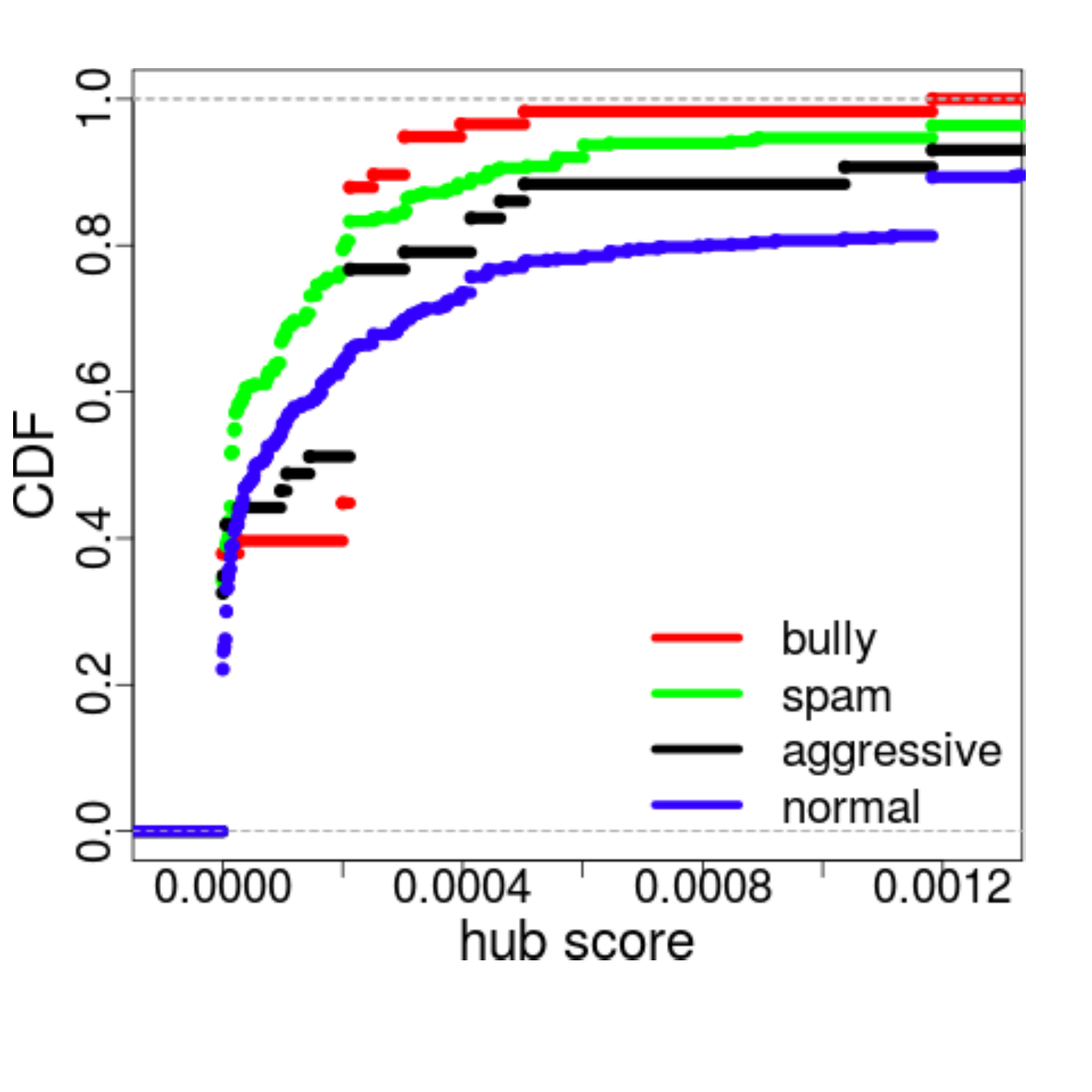}
		\vspace{-0.85cm}
		\caption{Hubs distribution.}
		\label{fig:ecdf-hubs}
	\end{subfigure}
	\begin{subfigure}[b]{0.21\textwidth}
		\captionsetup{font=scriptsize}
		\includegraphics[width=\textwidth]{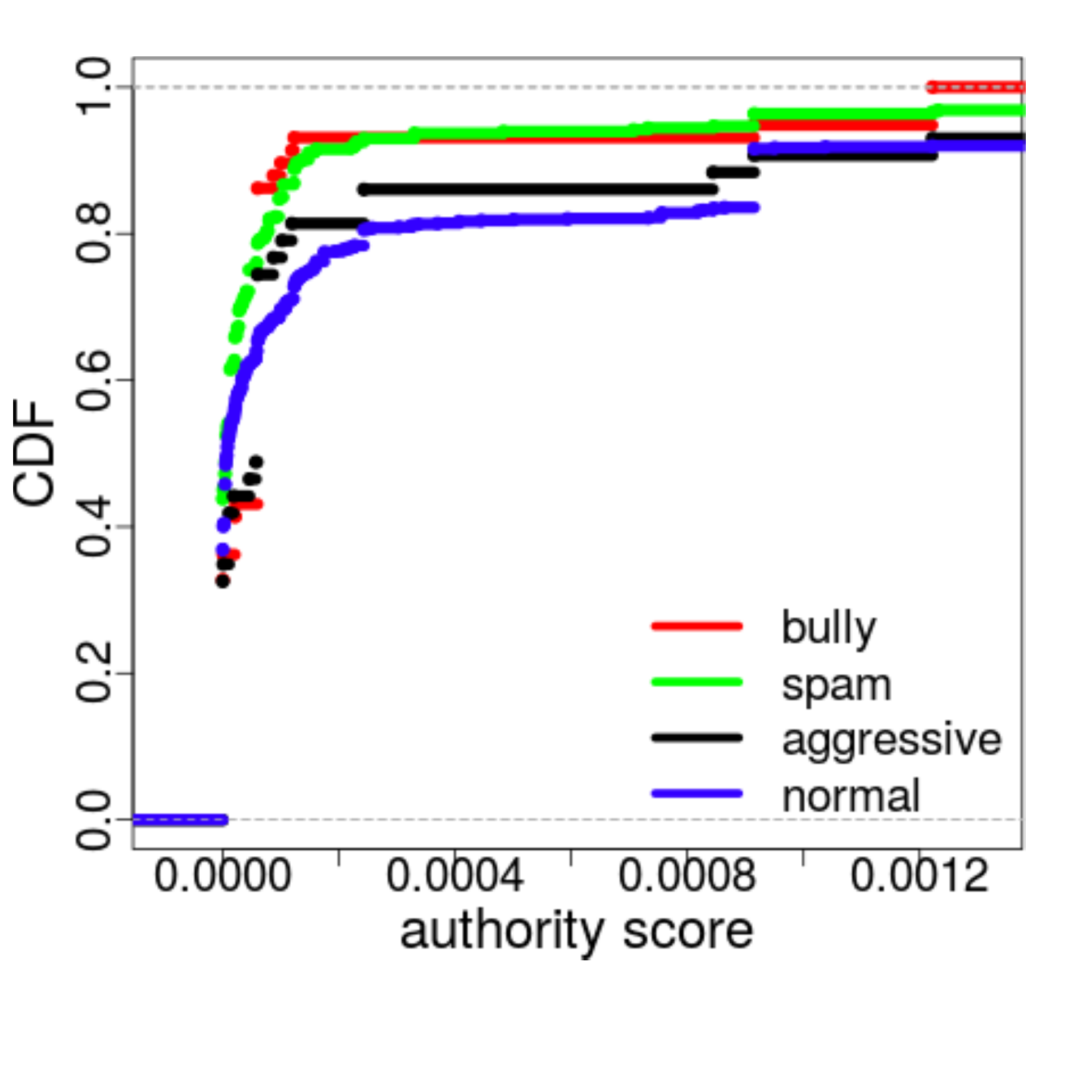}
		\vspace{-0.85cm}
		\caption{Authorities distribution.}
		\label{fig:ecdf-authorities}
	\end{subfigure}
	\begin{subfigure}[b]{0.21\textwidth}
		\captionsetup{font=scriptsize}
		\includegraphics[width=\textwidth]{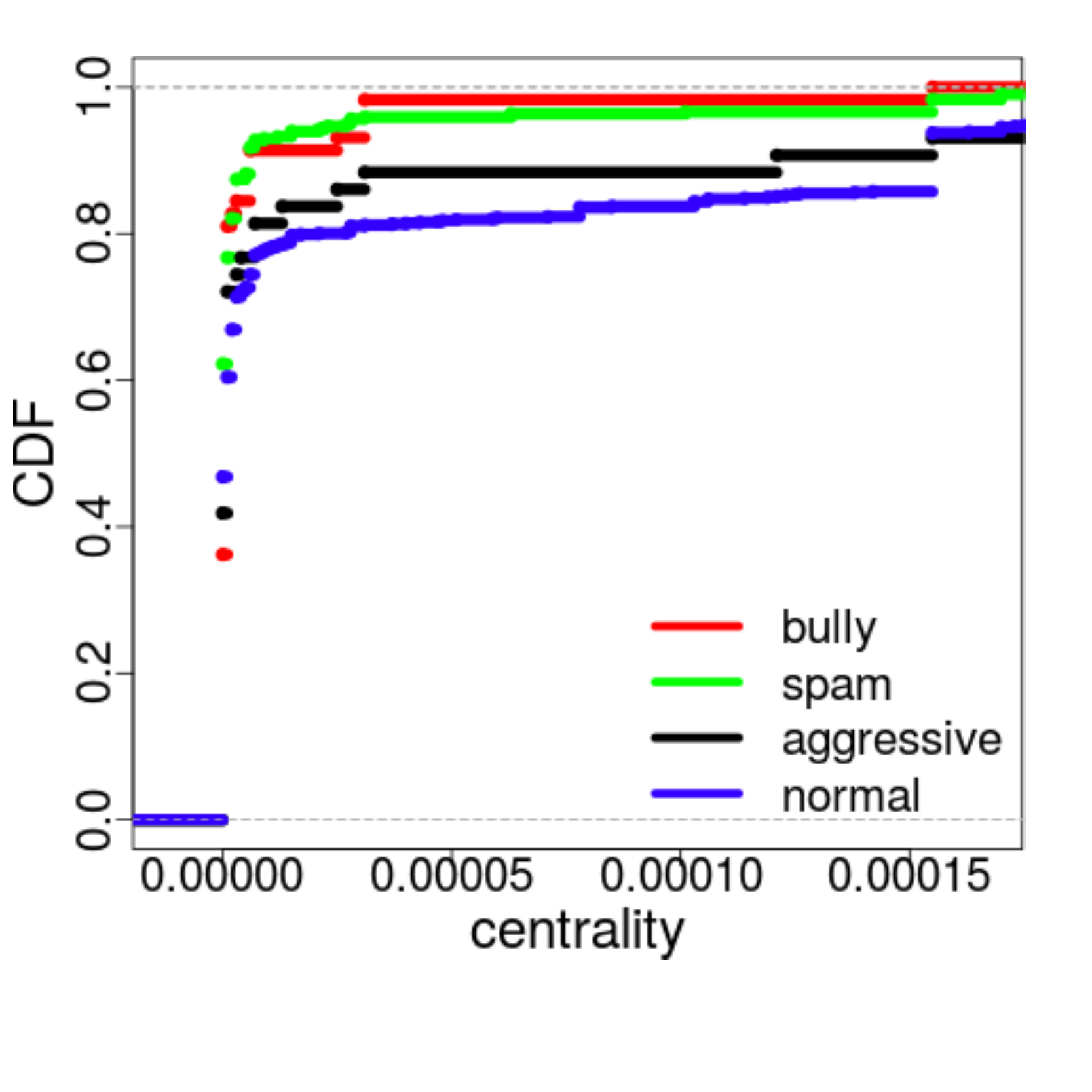}
		\vspace{-0.85cm}
		\caption{Eigenvector distribution.}
		\label{fig:ecdf-eigenvector}
	\end{subfigure}
	\begin{subfigure}[b]{0.21\textwidth}
		\captionsetup{font=scriptsize}
		\includegraphics[width=\textwidth]{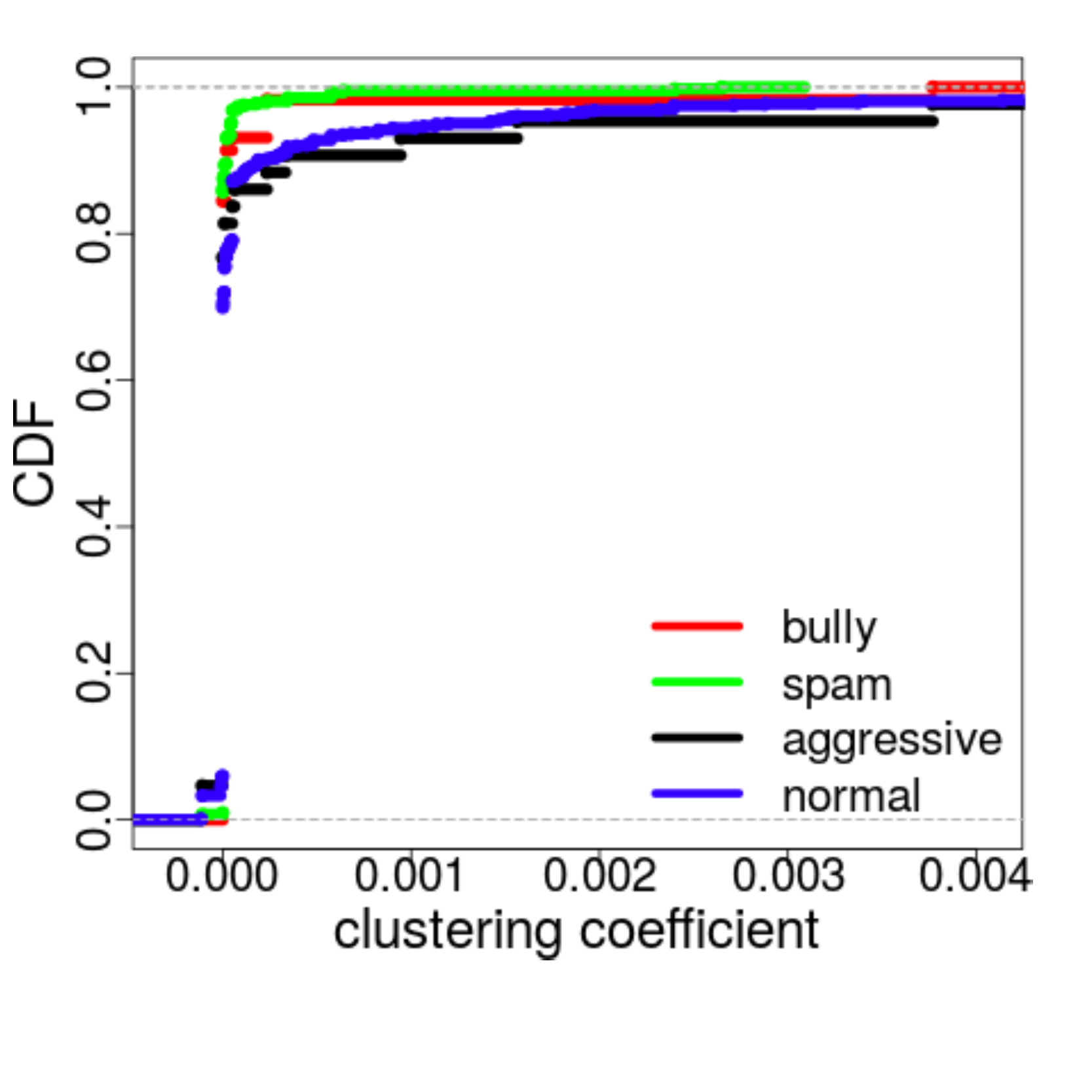}
		\vspace{-0.85cm}
		\caption{Clustering coef. distribution.}
		\label{fig:ecdf-clusteringCoef}
	\end{subfigure}
	\vspace{-0.2cm}
	\caption{Avg. distribution for (a) Hubs, (b) Authorities, (c) Eigenvectors, and (d) Clustering Coefficient.}
	\vspace{-0.4cm}
\end{figure}

\descr{Influence.}
Eigenvector centrality measures the influence of a user in his network, immediate or extended over multiple hops.
Closeness centrality measures the extent to which a user is close to each other user in the network.
To calculate the last four measures, we consider both the followers and friends relations of the users under examination in an undirected version of the network.
Figures~\ref{fig:ecdf-hubs},~\ref{fig:ecdf-authorities}, and~\ref{fig:ecdf-eigenvector} show the CDFs of the hubs (max value: $0.861$), authorities (max value: $0.377$), and eigenvector (max value: $0.0041$) scores for the four user classes.
We observe that bullies have lower values in their hub and authority scores which indicates that they are not so popular in their networks.
In terms of influence on their ego and extended network, they have behavior similar to spammers, while aggressors seem to have influence more similar to normal users.
We omit the CDF of the closeness centrality measure because we cannot reject the null hypothesis that the distributions are different.

\descr{Communities.}
Past work~\cite{hanish2004bullying} showed that bullies tend to experience social rejection from their environment and face difficulties in developing social relations.
We examine the usefulness of this attribute and calculate the clustering coefficient which shows a user's tendency to cluster with others.
Figure~\ref{fig:ecdf-clusteringCoef} plots the CDF of the clustering coefficient among the four classes.
We observe that bullies, similar to the spammers, are less prone to create clusters in contrast to aggressive and normal users.
Finally, we compute communities using the Louvain method~\cite{blondel2011louvain} which is suitable for identifying groups on large networks as it attempts to optimize the modularity measure (how densely connected the nodes within a cluster are) of a network by moving nodes from one cluster to another.
Overall, we observe a few communities with a high number of nodes (especially the network core) resulting in a feature which statistically differentiates bullies vs. normal users ($D$$=$$0.206$), but not aggressive vs. normal users ($D$$=$$0.1166$, $p$$=$$0.6364$).

\section{Modeling Aggressors~\&~Bullies}\label{sec:results} 

In this section, we present our efforts to model bullying and aggression behaviors on Twitter, using the features extracted and the labels provided by the crowdworkers as presented earlier.

\subsection{Experimental Setup}

We considered various machine learning algorithms, either probabilistic, tree-based, or ensemble classifiers (built on a set of classifiers whose individual decisions are then combined to classify data).
Due to limited space, we only present the best results with respect to training time and performance, obtained with the Random Forest classifier.
For all the experiments presented next, we use the WEKA data mining toolkit and repeated (10 times) 10-fold cross validation~\cite{Kim2009ErrorRate}, providing the relevant standard deviation (STD).
We do not balance the data to better match a real-world deployment.

\descr{Features selection.}
Most of the features presented in Section~\ref{sec:features} are useful (stat. significant) in discriminating between the user classes.
However, some are not useful and are excluded from the modeling analysis to avoid adding noise.
Specifically, we exclude the following features from our analysis: \emph{user-based} - verified account, default profile image, statistics on sessions, \emph{text-based} - average emotional scores, hate score, average word embedding, average curse score, and \emph{network-based} - closeness centrality and Louvain modularity.

\descr{Tree-based classifiers.}
Relatively fast compared to other classification models~\cite{Quinlan1986DecisionTrees}, tree classifiers have three types of nodes:
(i)~\textit{root} node, with no incoming edges, 
(ii)~\textit{internal} nodes, with one incoming edge and two or more outgoing edges,
(iii)~\textit{leaf} node, with incoming edge and no outgoing edges.
The root and each internal node correspond to feature test conditions (each test corresponds to a single feature) for separating data based on their characteristics, while leaf nodes correspond to the available classes.
We experimented with various tree classifiers: J48, LADTree, LMT, NBTree, Random Forest (RF), and Functional Tree; we achieved best performance with the RF classifier, which constructs a forest of decision trees with random subsets of features during the classification process.
To build the RF model, we tune the number of trees to be generated as 10, and the maximum depth unlimited.

\descr{Evaluation.}
For evaluation purposes, we examine standard machine learning performance metrics:
\emph{precision} (prec), \emph{recall} (rec), and weighted area under the ROC curve (AUC), at the class level and overall average across classes.
Also, the overall kappa (compares an observed accuracy with an expected accuracy, i.e., random chance), the root mean squared error (RMSE), which measures the differences among the values predicted by the model and the actually observed values, and finally the accuracy values are presented.

\descr{Experimentation phases.}
Two setups are tested to assess the feasibility of detecting user behavior:
(i) 4-classes classification: bully, aggressive, spam, and normal users,
(ii) 3-classes classification: bully, aggressive, and normal users.
This setup examines the case where we filter out spam with a more elaborate technique and attempt to detect the bullies and aggressors from normal users.

\subsection{Classification Results}

\descr{Detecting offensive classes.}
Here, we examine whether it is possible to distinguish between bully, aggressive, spam, and normal users.
Table~\ref{tbl:4_classes_bully_aggressive_spam_normal} overviews the results obtained with the RF classifier.
In more detail, we observe that the classifier succeeds to detect \textbf{43.2\%} (STD=0.042) of the bully cases, which is fair enough considering the small number of bully cases identified to begin with (only 4.5\% of our dataset).
In the aggressive case, we observe that recall is quite low, \textbf{11.8\%} (STD=0.078).
Based on the confusion matrix (omitted due to space limits), the misclassified cases mostly fall in either the normal or bullying classes, which aligns with the human annotations gathered during the crowdsourcing phase.
Overall, the average precision is \textbf{71.6\%}, and the recall is \textbf{73.32\%}, while the accuracy is \textbf{73.45\%}, with $0.4717$ kappa and $0.3086$ RMSE.

\begin{table}[!t]
    \begin{subtable}[b]{.5\linewidth}
      \captionsetup{font=scriptsize}
      \centering
      	\scalebox{0.7}{
        \begin{tabular}{l|lll}
        \hline                      & \textbf{Prec.} & \textbf{Rec.} & \textbf{AUC} \\ \hline
        bully          & 0.411          & 0.432         & 0.893        \\
        (STD)		   & 0.027			& 0.042         & 0.009        \\ \hline
        aggressive      & 0.295          & 0.118         & 0.793        \\
        (STD)		   & 0.054			& 0.078         & 0.036        \\ \hline
        spammer        & 0.686          & 0.561         & 0.808        \\
        (STD)		   & 0.008			& 0.010         & 0.002        \\ \hline
        normal         & 0.782          & 0.883         & 0.831        \\ 
        (STD.)		   & 0.004			& 0.005         & 0.003        \\ \hline \hline
        overall (avg.) & 0.718          & 0.733         & 0.815  \\     
        (STD)		   & 0.005			& 0.004         & 0.031        \\ \hline
        \hline
        \end{tabular}
        }
      \caption{4-classes classification.}            \label{tbl:4_classes_bully_aggressive_spam_normal}      
    \end{subtable}%
    \begin{subtable}[b]{.5\linewidth}
      \centering
       \captionsetup{font=scriptsize}
        \scalebox{0.7}{
        \begin{tabular}{l|lll}
        \hline
                                            & \textbf{Prec.} & \textbf{Rec.} & \textbf{AUC} \\ \hline
        \multicolumn{1}{l|}{bully}          & 0.555          & 0.609         & 0.912        \\
        					(STD)		    & 0.018			 & 0.029         & 0.009        \\ \hline
        \multicolumn{1}{l|}{aggressive}      & 0.304          & 0.114         & 0.812        \\
                			(STD)		    & 0.039			 & 0.012         & 0.015        \\ \hline
        \multicolumn{1}{l|}{normal}         & 0.951          & 0.976         & 0.911        \\ 
                			(STD)		    & 0.018			 & 0.029         & 0.009        \\ \hline\hline
        \multicolumn{1}{l|}{overall (avg.)} & 0.899          & 0.917         & 0.907        \\ 
                			(STD)		    & 0.016			 & 0.019         & 0.005        \\ \hline
        \end{tabular}
        }
               \caption{3-classes classification.}
       \label{tbl:3_classes_bully_aggressive_normal}
    \end{subtable} 
        \caption{Results on 4- and 3-classes classification.}
    \label{tbl:RF_results}    
    \vspace{-0.5cm}
\end{table}

\descr{Classifying after spam removal.}\label{subsec:3_class}
In this experimental phase, we want to explore whether the distinction between bully/aggressive and normal users will be more evident after applying a more sophisticated spam removal process in the preprocessing step.
To this end, we remove from our dataset all the cases identified by the annotators as spam, and re-run the RF modeling and classification.
Table~\ref{tbl:3_classes_bully_aggressive_normal} shows that, as expected, for bully cases there is an important increase in both the precision (\textbf{ 14.4\%}) and recall (\textbf{17.7\%}).
For aggressors, the precision and recall values are almost the same, indicating that further examination of this behavior is warranted in the future.
Overall, the average precision and recall of the RF model is \textbf{89.9\%} and \textbf{91.7\%}, respectively, while the accuracy is \textbf{91.08\%} with $0.5284$ kappa value and $0.2117$ RMSE.
Considering a \textbf{0.907} AUC, we believe that with a more sophisticated spam detection applied on the stream of tweets, our features and classification techniques can perform even better at detecting bullies and aggressors and distinguishing them from the typical Twitter users.

\descr{Discussion on AUC.}
ROC curves are typically used to evaluate the performance of a machine learning algorithm~\cite{davis2006relationship} by testing the system on different points and getting pairs of true positive (i.e., recall) against false positive rates indicating the sensitivity of the model.
The resulting area under the ROC curve can be read as the probability of a classifier correctly ranking a random positive case higher than a random negative case.
Our model performs well in both setups (Tables~\ref{tbl:4_classes_bully_aggressive_spam_normal} and~\ref{tbl:3_classes_bully_aggressive_normal}).
In the aggressor case, even though the recall value is low, the AUC is quite high because the false positive rate is especially low, with $0.008$ and $0.0135$ for the 4-class and 3-class classification, respectively.
We note that avoiding false positives is crucial to the successful deployment of any automated system aiming to deal with aggressive behavior.
Ideally, our classification system's results would be fed to a human-monitored system for thorough evaluation of the suspected users and final decision to suspend them or not, to reduce false positives.

\descr{Features evaluation.}
Table~\ref{tbl:features_evaluation} shows the top 12 features for each setup based on information gain.
Overall, in both setups the most contributing features tend to be user- and network-based, which describe the activity and connectivity of a user in the network.

\begin{table}[!t]
\begin{center}
\scalebox{0.7}{
\begin{tabular}{l|l}
\hline
{\bf Experiment}				&	{\bf Feature	(preserving order)}	\\
\hline
4-classes			& \#friends (11.43\%), reciprocity (11.10\%), \#followers (10.84\%)\\
					& \#followers/\#friends (9.82\%), interarrival (9.35\%), \#lists (9.11\%)\\
					& hubs (9.07\%), \#URLs (7.69\%), \#hashtags (7.02\%) \\
					& authority (6.33\%), account age (4.77\%), clustering coef. (3.47\%)\\
\hline
3-classes			& \#followers/\#friends (12.58\%), \#friends (12.56\%), \#followers (11.90\%)\\
					& interarrival (11.67\%), reciprocity (11.01\%), \#lists (9.57\%)\\
					& hubs (8.41\%), \#hashtags (6.2\%), \#posts (6.01\%)\\
					& account age (4.13\%), \#URLs (3.73\%), power difference (2.23\%)\\
\hline
\end{tabular}
}
\caption{Features evaluation.}
\label{tbl:features_evaluation}
\end{center}
\vspace{-0.2cm}
\end{table}

\begin{table}[!t]
      \centering
      \vspace{-0.3cm}
        \scalebox{0.75}{
        \begin{tabular}{l|lll}
        \hline
                                            & \textbf{Prec.} & \textbf{Rec.} & \textbf{ROC} \\ \hline
        \multicolumn{1}{l|}{bully}          & 1             & 0.667       & 0.833        \\
        \multicolumn{1}{l|}{aggressive}      & 0.5           & 0.4         & 0.757        \\
        \multicolumn{1}{l|}{normal}         & 0.931         & 0.971       & 0.82        \\ \hline
        \multicolumn{1}{l|}{overall (avg.)} & 0.909         & 0.913       & 0.817        \\ \hline
        \end{tabular}
        }
             \caption{Classification on balanced data.}
       \label{tbl:3_class_results_balanced}
       \vspace{-0.15cm}
\end{table}

\descr{Balancing data.}
Based on~\cite{chen2004using}, and similar to almost all classifiers, Random Forest suffers from appropriately handling extremely imbalanced training dataset (similar to our case) resulting in bias towards the majority classes.
To address this issue, we over-sample (based on SMOTE~\cite{Chawla2002}, which creates synthetic instances of the minority class) and under-sample (a resampling technique without replacement) at the same time, as it has proven to result in better overall performance~\cite{Chawla2002}.
Here, we focus on the 3-class experimentation setup, i.e., without considering the spam user class.
After randomly splitting the data into 90\% for training and 10\% for testing sets, we proceed with the balancing of the training set.
The resulting data distribution is 349, 386, and 340 instances for the bully, aggressive, and normal classes, respectively.
We note there is no resampling of the test set.
Table~\ref{tbl:3_class_results_balanced} shows the classification results.
After balancing the data, the classifier detects \textbf{66.7\%} and \textbf{40\%} of the bully and aggressive cases, respectively, while overall, the accuracy is $91.25\%$, with $0.5965$ kappa and $0.1423$ RMSE.

\subsection{Twitter Reaction to Aggression}

Recently, Twitter has received a lot of attention due to the increasing occurrence of harassment incidents~\cite{guardiantrolls}.
While some shortcomings have been directly acknowledged by Twitter~\cite{guardianfail}, they do act in some cases.
To understand the impact of our findings, we make an estimate of Twitter's current effectiveness to deal with harassment by looking at account statuses. 
Twitter accounts can be in one of three states: \emph{active}, \emph{deleted}, or \emph{suspended}.
Typically, Twitter suspends an account (temporarily or even permanently) if it has been compromised, is considered spam/fake, or if it is \emph{abusive}~\cite{twitterSuspendedAccounts}.

\begin{table}[!t]
\vspace{-0.3cm}
    \begin{subtable}[b]{.5\linewidth}
      \captionsetup{font=scriptsize}
      \centering
      	\scalebox{0.7}{
			\begin{tabular}{@{}llll@{}}
			\toprule
			          & \textbf{active} & \textbf{deleted} & \textbf{suspended} \\ \midrule
			bully     & 67.24\%         & 32.76\%          & 0.00\%             \\
			aggressive & 65.12\%         & 20.93\%          & 13.95\%            \\
			normal    & 86.53\%         & 5.72\%           & 7.75\%             \\
			 \bottomrule
			\end{tabular}
        }
      \caption{Status check on Nov 2016.}
      \label{tbl:status_check_nov}              
    \end{subtable}%
    \begin{subtable}[b]{.5\linewidth}
      \centering
       \captionsetup{font=scriptsize}
        \scalebox{0.7}{
        \begin{tabular}{@{}llll@{}}
        \toprule
                  & \textbf{active} & \textbf{deleted} & \textbf{suspended} \\ \midrule
        bully     & 62.07\%         & 37.93\%          & 0.00\%            \\
        aggressive & 55.81\%         & 25.58\%          & 18.60\%            \\
        normal    & 85.01\%         & 6.86\%           & 8.13\%                \\ \bottomrule
        \end{tabular}
        }
       \caption{Status check on Feb 2017.}
        \label{tbl:status_check_feb}        
    \end{subtable} 
    \vspace{-0.5cm}
\caption{Distribution of users' behaviors in twitter statuses.}
    \label{tbl:twitter_status}    
    \vspace{-0.4cm}
\end{table}

After our initial experiments, we checked the current status of all labeled users in our dataset.
The status check was performed over two time periods: at the end of November 2016 and February 2017.
Tables~\ref{tbl:status_check_nov},~\ref{tbl:status_check_feb} show the break down of account statuses for each label for the two time periods.
From the more recent time period (February 2017), we observe that the majority of ``bad'' users in our dataset have suffered no consequences from Twitter: $55.81\%$ of aggressive and $62.07\%$ of cyberbullying accounts were still active.
In particular, suspension (Twitter-taken action) and deletion (the user removing his account) statuses exhibit a stark contrast.

While $18.6\%$ of aggressive users are suspended, {\em no} cyberbullying users are.
Instead, bullies tend to delete their accounts proactively ($\sim$$38\%$).
Comparing the statuses of aggressive users between the two time periods, we see an  increase ($4.65\%$) in the percentage of those suspended.
This is in alignment with Twitter's recent efforts to combat harassment cases, for instance, by preventing suspended users from creating new accounts~\cite{CNNtech}, or temporarily limiting users for abusive behavior~\cite{independent}.
However, in both time periods, there are {\em no} suspended bully users.
Again, bully users seem to delete their accounts, perhaps in an attempt to prevent suspension.
Regardless of the reason, for these users we also observe a $5.17\%$ increase in deleted accounts between the two time periods.
The lack of suspension of bully users could because bullying often manifests in a hidden fashion, e.g., within seemingly innocuous criticisms, yet are repetitive (and thus harmful over time) in nature~\cite{mcmahon2000bullying}.

\section{Discussion \& Conclusion}\label{sec:discussion}
Although the digital revolution and the rise of social media enabled great advances in communication platforms and social interactions, wider proliferation of harmful behavior has also emerged.
Unfortunately, effective tools for detecting harmful actions are scarce, as this type of behavior is often ambiguous in nature and/or exhibited via seemingly superficial comments and criticisms.
Aiming to address this gap, this paper presented a novel system geared to automatically classify two kinds of harmful online behavior, cyberaggression and cyberbullying, focusing on Twitter.

We relied on crowd-workers to label $1.5k$ users as normal, spammers, aggressive, or bullies, from a corpus of $\sim$$10k$ tweets (distilled from a larger set of $1.6M$ tweets), using an efficient, streamlined labeling process.
We investigated $30$ features from $3$ types of attributes (user, text, network based) characterizing such behavior.
We found that bullies are less popular (fewer followers/friends, lower hub, authority, eigenvector scores) and participate in few communities.
Although they are not very active as per number of posts overall, when they do become active, they post more frequently than typical users, and do so with more hashtags, URLs, etc.
Based on the analysis in~\cite{chatzakou2017hypertext} they also are older Twitter users based on their account age.

Aggressive users show similar behavior to spammers in terms of number of followers, friends, and hub scores.
Similar to bullies, they also do not post a lot of tweets, but exhibit a small response time between postings, and often use hashtags and URLs in their tweets.
They also tend to have been on Twitter for a long time, like bullies.
However, their posts seem to be more negative in sentiment than bullies or normal users.
On the other hand, normal users are quite popular with respect to number of followers, friends, hubs, authorities.
They participate in many topical lists, and use few hashtags and URLs.
These observations are in line with the intuition that bully and aggressive users tend to attack, in rapid fashion, particular users or groups they target, and do so in short bursts, with not enough duration or content to be detected by Twitter's automated systems.
In general, we find that aggressive users are more difficult to characterize and identify using a machine learning classifier than bullies, since sometimes they behave like bullies, but other times as normal or spam users.

We showed that our methodology for data analysis, labeling, and classification can scale up to millions of tweets, while our machine learning model built with a Random Forest classifier can distinguish between normal, aggressive, and cyberbullying users with high accuracy ($>$ $91\%$).
While prior work almost exclusively focused on user- and text-based features (e.g., linguistics, sentiment, membership duration), we performed a thorough analysis of network-based features, and found them to be very useful, as they actually
are the most effective for classifying aggressive user behavior (half of the top-$12$ features in discriminatory power are network-based).
Text-based features, somewhat surprisingly, do not contribute as much to the detection of aggression (with an exception of tweet characteristics, such as number of URLs, hashtags, and sentiment).

Finally, we discussed the effectiveness of our detection method by comparing prediction results of the examined users with the suspension and deletion of their accounts as observed in the wild.
We found that bullies tend not to have been suspended, but instead, take seemingly proactive measures and delete their accounts.
Contrary, aggressors are suspended more often than bullies or normal users, which is in line with recent Twitter efforts to combat harassment by preventing suspended users from creating new accounts~\cite{CNNtech} or temporarily limiting users for abusive behavior~\cite{independent}.

\descr{Acknowledgements.} This research has been funded by the European Commission as part of the ENCASE project  (H2020-MSCA-RISE of the European Union under GA number 691025) and by the EPSRC under grant number N008448. We would also like to thank Andri Ioannou for her feedback throughout this project.

\bibliographystyle{abbrv}

\end{document}